\documentclass[prd,twocolumn,superscriptaddress,altaffilletter,amssymb,showpacs,nofootinbib]{revtex4}
\usepackage[dvips]{graphicx}
\usepackage{amsmath}
\newcommand{\be}{\begin{equation}}
\newcommand{\ee}{\end{equation}}
\newcommand{\bea}{\begin{eqnarray}}
\newcommand{\eea}{\end{eqnarray}}
\newcommand{\vphi}{\varphi}

\begin{document}

\title{Does Dirac-Born-Infeld Modification Of Quadratic Theories Really Matter?}

\author{Israel Quiros}\email{iquiros@fisica.ugto.mx}\affiliation{Divisi\'on de Ciencias e Ingenier\'ia de la Universidad de Guanajuato, A.P. 150, 37150, Le\'on, Guanajuato, M\'exico.}

\author{L. Arturo Ure\~na-L\'opez}\email{lurena@fisica.ugto.mx}\affiliation{Divisi\'on de Ciencias e Ingenier\'ia de la Universidad de Guanajuato, A.P. 150, 37150, Le\'on, Guanajuato, M\'exico.}

\date{\today}

\begin{abstract}
We study the consequences of further modification of $f(R,R_{\mu\nu}R^{\mu\nu},R_{\mu\nu\sigma\rho}R^{\mu\nu\sigma\rho})/f(R)$-theories by means of the Dirac-Born-Infeld deformation procedure, which amounts to the replacement of $f$ by $\lambda(\sqrt{1+2f/\lambda}-1)$ (the free parameter $\lambda$ fixes an additional energy scale). We pay special attention to the definition of masses of the linearized propagating degrees of freedom since these are important to judge about the stability of the linearization around vacuum background spaces. In this context we discuss the subtleties associated with expanding $f(R,R_{\mu\nu}R^{\mu\nu},R_{\mu\nu\sigma\rho}R^{\mu\nu\sigma\rho})$-Lagrangians around maximally symmetric spaces of constant curvature, as well as with equivalence of the linearized Lagrangian to a scalar-tensor theory. Investigation of the consequences of applying the Dirac-Born-Infeld strategy to further modify quadratic theories, on the stability of de Sitter vacuum, as well as its impact on the cosmological dynamics, is the main concern of this paper. 
\end{abstract}


\maketitle

\section{Introduction}

Attempts to modify the Einstein-Hilbert (EH) action of general relativity (GR) $$S_{EH}=\frac{1}{2\kappa^2}\int d^4x\sqrt{|g|}\left(R-2\Lambda\right),$$ where $R=g^{\mu\nu}R_{\mu\nu}$ is the Ricci curvature scalar, and $\Lambda$- the cosmological constant ($\kappa^2=m_{Pl}^{-2}=8\pi G$), have been motivated by a number of reasons. In particular, renormalization at one-loop demands that the Einstein-Hilbert action be supplemented by higher order curvature terms \cite{udw}.\footnote{Higher order actions are indeed renormalizable (but not unitary) \cite{stelle1}).} Besides, when quantum corrections or string theory are taken into account, the effective low energy action for pure gravity admits higher order curvature invariants \cite{qstring}. 

More recently it has been suggested that the present cosmic speed-up
could have its origin in -- among other possibilities -- corrections
to the GR equations of motion, generated by non-linear contributions
of the scalar curvature $R$ in the pure gravity Lagrangian of $f(R)$
theories \cite{odintsov1,frspeedup,positive,carroll}. Solar system
constraints on $f(R)$ theories that are able to accommodate present
accelerated expansion of the Universe, have been one of the most
discussed subjects lately
\cite{olmo2005,olmoprd2007,odintsovrev,faraonirev,faraoni2008,sotiriou2008},
see also\cite{DeFelice:2010aj} for an extensive review. Comparison of these theories with solar system measurements, relies on the weak-field limit expansion of the $f(R)$ Lagrangian, and the consequent calculation of post-Newtonian contributions to the metric coefficients \cite{olmoprd2007,faraonirev,olmo2005}. Nonetheless, even if $f(R)$ theories were not a viable alternative to explain current acceleration of the expansion, their relevance to study early-time inflation \cite{starobinsky} might fuel further interest in these alternatives to general relativity. 

Next in degree of complexity are the so called $f(R,R_{\mu\nu}R^{\mu\nu},R_{\mu\nu\sigma\rho}R^{\mu\nu\sigma\rho})$ -- $f(R,...)$ for short -- theories \cite{carroll,stelle,ovrut,navarro,solganik,chiba}. The gravitational spectrum of the linearization of these theories consists of a massless spin-2 graviton plus two additional gravitational propagating degrees of freedom: a massive spin-0 excitation, and a massive spin-2 propagating mode. The latter appears to be a ghost mode associated with the Weyl curvature invariant $C^2\equiv C_{\mu\nu\sigma\lambda}C^{\mu\nu\sigma\lambda}$ \cite{stelle,ovrut,solganik}. Notwithstanding, there are ways to overcome (or at least to smooth out) the consequences of the would be massive spin-2 ghost mode \cite{navarro}.

There are additional ways to modify the EH GR action. For instance, the one based on the Dirac-Born-Infeld (DBI) procedure for smoothing out singularities \cite{carroll,dbiproc,fiorini}.\footnote{The proposal to remove initial as well as final singularities in modified gravity has been given in Ref. \cite{odi}. It was shown there that the addition of a $R^2$ term to otherwise divergent modified gravity makes it regular.} According to this procedure the original Lagrangian density ${\cal L}=\sqrt{|g|}L$ -- whose singularities are to be cured -- is replaced by one of the DBI form: $${\cal L}\;\rightarrow\;{\cal L}_{DBI}=\sqrt{|g|}\lambda\left(\sqrt{1+\frac{2L}{\lambda}}-1\right),$$ where the scale $\lambda$ sets an upper bound to curvatures accessible to the theory. A combination of the above possible modifications, i. e., a DBI-type action containing an $f(R,...)$ function within the square root, i. e., a replacement of 

\be f(R,...)\rightarrow\lambda\left(\sqrt{1+\frac{2f(R,...)}{\lambda}}-1\right)\;,\label{dbi modification}\ee in the action $S\propto\int d^4x\sqrt{|g|}f(R,...)$, could supply an additional cosmological scenario where to look for alternative explanations to several phenomena such as inflation and the present speedup of the cosmic expansion. Several theories of gravity of this kind have been proposed since long ago in \cite{eddington}, and in more recent years, for instance, in \cite{deser} (see also \cite{comelli}). To be phenomenologically viable, non-linear modifications of general relativity have to satisfy several physically motivated requirements \cite{deser}: i) reduction to EH action at small curvature, ii) ghost freedom, iii) regularization of some singularities (as, for instance, the Coulomb-like Schwarzschild singularity), and iv) supersymmetrizability. Nonetheless, the latter requirement is quite stringent and, for most purposes, might be excluded.

Would further modification of $f(R,....)/f(R)$ theories of gravity, through the DBI deformation strategy (\ref{dbi modification}), help surmounting the severe problems related with the presence of a multitude of instabilities within these theories? Would it modify the asymptotic properties of the cosmic dynamics? Aim of the present paper is, precisely, to investigate the consequences of applying the Dirac-Born-Infeld procedure to $f(R,...)$ theories of gravity -- including $f(R)$ gravity as a particular case --, regarding stability of de Sitter vacuum solutions, as well as its impact on the cosmological dynamics. 

The paper has been organized as follows. In section II the subtleties associated with expanding $f(R,...)$ gravity theories about maximally symmetric spaces of constant curvature are discussed. This topic is central to judge about stability of the propagating degrees of freedom upon linearization. The equivalence of ghost-free $f(R,...)$ gravity to a scalar-tensor theory is demonstrated in the same section. Sections III and IV are devoted to study modifications of the stability properties of $f(R,...)$, and $f(R)$ theories, respectively, after applying to them the DBI procedure. The consequences for the cosmological dynamics of DBI-modified $f(R,...)/f(R)$ gravity is the main concern of section V. The results of the present investigation are discussed in section VI, while the conclusions are given in section VII. An appendix with brief and concise tips on how to apply the dynamical systems tools, is included in the final section VIII.

\section{Expansion around maximally symmetric spaces of constant curvature}

Here we try exposing the importance of considering a consistent expansion around maximally symmetric vacuum spaces of constant curvature of higher order $f(R,...)$ theories, to judge about stability issues. The results obtained can be applied also to $f(R)$ theory as a particular case. In what follows, for simplicity of writing we use the following definition of variables $X^i=(X,Y,Z)$:

\be X\equiv R\;,\;\;Y\equiv R_{\mu\nu}R^{\mu\nu}\;,\;\;Z\equiv R_{\mu\nu\sigma\lambda}R^{\mu\nu\sigma\lambda}\;,\label{variables}\ee where $R$ is the curvature scalar, $R_{\mu\nu}$ - the Ricci tensor, and $R_{\mu\nu\sigma\lambda}$ - the Riemann tensor. In terms of these variables the Gauss-Bonnet invariant ${\cal G}$ and the Weyl invariant $C^2\equiv C_{\mu\nu\varphi\lambda}C^{\mu\nu\varphi\lambda}$ can be written as

\be {\cal G}=X^2-4Y+Z\;,\;\;\text{and}\;\;C^2=Z-2Y+\frac{1}{3}X^2\;,\label{invariants}\ee respectively. Combining these expressions one can get the following equalities that will be useful latter on:

\be Y=\frac{1}{2}C^2-\frac{1}{2}{\cal G}+\frac{1}{3}X^2\;,\;\;Z=2C^2-{\cal G}+\frac{1}{3}X^2\;.\label{yz}\ee

We will consider pure gravitational actions of the following kind:

\be S_g=\frac{1}{2\kappa^2}\int dx^4\sqrt{|g|}\;f(X,Y,Z)\;,\label{action}\ee where $\kappa^2=8\pi G_N=M_{Pl}^{-2}$. The following field equations can be obtained from the above action by varying with respect to the metric $g_{\mu\nu}$ \cite{carroll}:

\bea f_X G_{\mu\nu}=\frac{1}{2}g_{\mu\nu}(f-X f_X)-(g_{\mu\nu}\Box-\nabla_\mu\nabla_\nu) f_X\nonumber\\
-2(f_Y R_{\;\;\mu}^{\sigma}R_{\sigma\nu}+f_Z R_{\lambda\sigma\rho\mu}R^{\lambda\sigma\rho}_{\;\;\;\;\;\;\nu})-g_{\mu\nu}\nabla_\sigma\nabla_\lambda(f_Y R^{\sigma\lambda})\nonumber\\-\Box(f_Y R_{\mu\nu})+2\nabla_\sigma\nabla_\lambda(f_Y R^\sigma_{\;\;(\mu}R^\lambda_{\;\;\nu)}+2f_Z R^{\sigma\;\;\;\;\;\lambda}_{\;(\mu\nu)})\;,\label{feqs}\eea where, as usual, $G_{\mu\nu}\equiv R_{\mu\nu}-g_{\mu\nu}R/2$ is the Einstein tensor, $\Box\equiv g^{\mu\nu}\nabla_\mu\nabla_\nu$ is the D'Lambertian, and we have used the usual representation for symmetrization: $T_{(\mu\nu)}=(T_{\mu\nu}+T_{\nu\mu})/2$. In the presence of matter these field equations amount to:

\be G_{\mu\nu}=8\pi G_N^{eff}(T_{\mu\nu}^{(m)}+T_{\mu\nu}^{cur})\;,\label{full feqs}\ee where $8\pi G_N^{eff}=\kappa^2/f_X$ is the effective gravitational coupling, $T_{\mu\nu}^{(m)}$ is the stress-energy tensor for matter, while $T_{\mu\nu}^{cur}$ equals $\kappa^{-2}$ times the right-hand-side (RHS) of equation (\ref{feqs}). The trace of equation (\ref{feqs}) generates an additional constraint on the curvature:

\bea &&2f-Xf_X-2Yf_Y-2Zf_Z\nonumber\\
&&\;\;\;-\Box(3f_X+2Xf_Y+Xf_Z+\frac{1}{8}X^2f_Y)=\kappa^2 T^{(m)}\;.\label{trace}\eea

Let us consider expanding the above action (\ref{action}) around maximally symmetric vacuum spaces of constant curvature $R=R_0$, i. e., we will Taylor expand $f(X,Y,Z)$ in the neighborhood of the point $(X_0,Y_0,Z_0)$ where $X_0=R_0$, $Y_0=R_0^2/4$ and $Z_0=R_0^2/6$, and $f(X_0,Y_0,Z_0)=f_0$, up to the second order $\sim (X-X_0)^2\sim(Y-Y_0)^2\sim(Z-Z_0)^2$. One has:

\bea &&f(X,Y,Z)=f_0+\sum_i \left(\frac{\partial f}{\partial X^i}\right)_0 (X^i-X_0^i)\nonumber\\&&+\frac{1}{2}\left(\frac{\partial^2 f}{\partial X^i \partial X^k}\right)_0 (X^i-X^i_0)(X^k-X^k_0)+{\cal O}(3)\;,\label{taylor}\eea or, in explicit form:

\bea &&f(X,Y,Z)=f_0+f_X^0 (X-X_0)+f_Y^0 (Y-Y_0)\nonumber\\ &&\;\;\;\;\;\;\;\;\;\;\;\;\;\;\;\;\;\;\;\;\;\;\;\;\;\;+f_Z^0 (Z-Z_0)+\frac{1}{2}f_{XX}^0 (X-X_0)^2\nonumber\\&&+f_{YX}^0(X-X_0)(Y-Y_0)+f_{ZX}^0(X-X_0)(Z-Z_0)\nonumber\\&&\;\;\;\;\;\;\;\;\;+\frac{1}{2}f_{YY}^0(Y-Y_0)^2+f_{ZY}^0(Y-Y_0)(Z-Z_0)\nonumber\\&&\;\;\;\;\;\;\;\;\;\;\;\;\;\;\;\;\;\;\;\;\;\;\;\;\;\;\;\;\;\;\;\;+\frac{1}{2}f_{ZZ}^0(Z-Z_0)^2+{\cal O}(3)\;,\label{taylor explicit}\eea where $$f_X^0\equiv\left(\frac{\partial f}{\partial X}\right)_0\;,\;\;f_{XX}^0\equiv\left(\frac{\partial^2 f}{\partial X\partial X}\right)_0,$$ etc. Notice that, while keeping up to the second order in the expansion is legitimate since we are considering small $\delta X^i=X^i-X_0^i\ll 1$, considering up to quadratic terms in the curvature $\sim R^2$ is an additional requirement that has nothing to do with the order of the perturbations around $X_0^i=(X_0,Y_0,Z_0)$ one is being considering. Actually, the expansion (\ref{taylor},\ref{taylor explicit}) does not exclude large curvature $R_0$. The only requirement is that if $X_0=R_0$ is large, so is $X$. However, in this latter case it makes sense to keep only the terms with the higher orders in the curvature (terms $\propto X^4, Y^2, Z^2$).

After had clarified this point let us consider small curvature backgrounds only, i. e., small $R_0$, so that it makes sense to keep terms up to the order $\sim{\cal O}(R^2)$ in (\ref{taylor explicit}):

\bea &&f(X,Y,Z)=f_0-X_0 f_X^0-Y_0 f_Y^0-Z_0 f_Z^0\nonumber\\&&\;\;\;\;\;\;\;\;\;\;\;\;\;\;\;\;\;\;+\frac{1}{2}X_0^2 f_{XX}^0+(f_X^0-X_0 f_{XX}^0) X\nonumber\\&&\;\;\;\;\;\;\;\;\;\;\;\;\;\;\;\;+\frac{1}{2}f_{XX}^0 X^2+f_Y^0 Y+f_Z^0 Z+{\cal O}(R^3)\;.\eea 

By using the equations (\ref{yz}) one can remove the variables $Y$, $Z$ from the expansion, so that, finally:

\be f(X,Y,Z)=\lambda_0+\alpha_0 X+\frac{\beta_0}{6}X^2+\frac{\gamma_0}{2} C^2\;,\label{final}\ee where

\bea \lambda_0\equiv f_0-X_0 f_X^0-Y_0 f_Y^0-Z_0 f_Z^0+\frac{X_0^2}{2}f_{XX}^0\;,\nonumber\\
\alpha_0\equiv f_X^0-X_0 f_{XX}^0\;,\;\;\beta_0\equiv3f_{XX}^0+2f_Y^0+2f_Z^0\;\nonumber\\
\gamma_0\equiv f_Y^0+4f_Z^0\;.\label{coefficients}\eea 

As a check of consistency, notice that by taking into account the trace equation (\ref{trace}), which in the present situation amounts to:

\be 2f_0-X_0f_X^0-2Y_0f_Y^0-2Z_0f_Z^0=0\;,\label{trace'}\ee then the coefficient $$\lambda_0=-\frac{X_0}{2}(f_X^0-X_0f_{XX}^0)=-\frac{\alpha_0}{2}X_0\;.$$ This means that the definition of the cosmological constant $\Lambda=-\lambda_0/2\alpha_0$ (see below) leads to the expected result $\Lambda=X_0/4$.

As customary one can introduce, additionally, the following magnitudes:

\bea m_0^2=\frac{\alpha_0}{\beta_0}=\frac{f_X^0-X_0 f_{XX}^0}{3f_{XX}^0+2f_Y^0+2f_Z^0}\;,\nonumber\\
m_2^2=-\frac{\alpha_0}{\gamma_0}=-\frac{f_X^0-X_0 f_{XX}^0}{f_Y^0+4f_Z^0}\;,\;\;-2\Lambda=\frac{\lambda_0}{\alpha_0}\;,\label{masses}\eea so that (\ref{final}) can be written in the following form:

\be f(X,Y,Z)=\alpha_0\left(-2\Lambda+R+\frac{1}{6m_0^2}R^2-\frac{1}{2m_2^2}C^2\right)\;.\label{linearized}\ee The gravitational spectrum of the corresponding linearized theory consists of a standard massless spin-2 excitation, plus a spin-0 mode of mass squared $m_0^2$, and an additional spin-2 mode of mass squared $m_2^2$ \cite{stelle,ovrut} -- $m_0^2$ and $m_2^2$ are given by (\ref{masses}). The latter spin-2 excitation occurs to be a ghost, leading to non-unitary states upon quantization (otherwise unitary states of negative energy). \footnote{In Ref. \cite{solganik} this has been shown starting directly from a general action of the form (\ref{action}) by investigating the propagator of the linearized degrees of freedom.}

\subsection{Stability Requirements}

Stability issues are central in the study of higher order modifications of general relativity, since these are plagued by several kinds of instabilities, some of which are catastrophic, leading to subsequent ruling out of the corresponding theories. Amongst these is the fundamental Ostrogradski instability, based on the powerful no-go theorem of the same name \cite{woodard}: ``there is a linear instability in the Hamiltonians associated with Lagrangians which depend upon more than one time derivative in such a way that the dependence cannot be eliminated by partial integration''. This result is general and can be extended to higher order derivatives in general. As a consequence, the only Ostrogradski-stable higher order modifications of Einstein-Hilbert action are those in the form of an $f(X)\equiv f(R)$ function \cite{woodard}. This result alone might rule out any intent to consider higher order modifications such as quadratic ones, for instance. However, the subject is subtle and, in the last instance, consideration of such theories can shed more light on the stability issue. As an example, consider theories of the kind $f(X,Y,Z)$ where the invariants $Y$, $Z$, enter in the same combination as in the Gauss-Bonnet invariant ${\cal G}$, i. e., in the combination $-4Y+Z$. In this case, upon linearizing around maximally symmetric spaces of constant curvature, since $f_Y^0+4f_Z^0=0$, the massive propagating spin-2 ghost associated with the Weyl invariant decouples from the gravitational spectrum. Besides, the linearized theory can be recast into the form of an $f(X)$-theory (see Eq. (\ref{linearized})), so that it is also Ostrogradski-stable.

Another kind of catastrophic instability is the so called ``Ricci instability'' \cite{faraonirev}, also known as Dolgov-Kawasaki instability \cite{kawasaki}. For arbitrary $f(X)$-theories the analysis of Ricci stability has been generalized in \cite{faraoni2006}, while a consistent and simple physical interpretation has been given by the same author in Ref. \cite{faraoni2007}. According to the latter interpretation, assuming that the effective gravitational coupling $8\pi G_N^{eff}\equiv\kappa^2/f_X$ (see Eq. (\ref{full feqs})) is positive, then, since $$\frac{dG_N^{eff}}{dX}=-\frac{\kappa^2 f_{XX}}{8\pi (f_X)^2}\;,$$ for negative $f_{XX}<0$, the effective gravitational coupling $G_N^{eff}$ increases with the curvature, otherwise, at large curvature gravity becomes stronger, and since $X$ itself generates larger and larger curvature through the trace equation Eq. (\ref{trace}), the effect becomes uncontrollably stronger because of an increased $G_N^{eff}$. In other words, a positive feedback mechanism acts to destabilize the theory \cite{faraonirev}. Hence, to avoid the ``Ricci instability'' it is necessary that $f_{XX}\geq 0$. 

Other instabilities, such as those caused by the presence of a spin-0 tachyon degree of freedom are not of less importance. The latter instability is associated with negative values of the spin-0 mass squared $m_0^2$ in Eq. (\ref{masses}). Last but not least, there is an additional requirement that has not been discussed in detail in the literature. As seen from (\ref{linearized}), the constant $\alpha_0\equiv f_X^0-X_0 f_{XX}^0$ is an overall factor that multiplies the linearized action (\ref{action}), and, hence, it may change the sign of the action, otherwise, the sign of the effective gravitational coupling upon linearization. Therefore, if one considers $f(X)$ theories; Ricci stability, absence of tachyon, and positivity of the effective gravitational coupling, are not independent requirements. Actually, in this case $$m_0^2=\frac{f_X^0-X_0 f_{XX}^0}{3f_{XX}^0}\;,$$ so that, Ricci stability ($f^0_{XX}>0$), and positivity of the effective gravitational coupling upon linearization ($f_X^0-X_0 f_{XX}^0>0$), together imply absence of tachyon instability $m_0^2>0$. For arbitrary $f(X,Y,Z)$ the above requirements are independent.

To summarize the discussion on the relevant stability requirements imposed on higher order curvature modifications of Einstein-Hilbert theory, here we list them:

\begin{itemize}

\item{Ostrogradski Stability:} The linearized $f(X,Y,Z)$ theory should be expressible as an equivalent $f(X)$-theory.\footnote{Stated in this form Ostrogradski stability implies also, absence of spin-2 Weyl ghost propagating modes.} This can be implemented if $$f_Y^0+4f_Z^0=0\;.$$

\item{Ricci Stability:} $$f_{XX}^0\geq 0\;.$$

\item{Absence of Tachyon Instability:} $$m_0^2=\frac{f_X^0-X_0 f_{XX}^0}{3f_{XX}^0+2f_Y^0+2f_Z^0}\geq 0\;.$$

\item{Non-negativity of the Effective Gravitational Coupling:} $$\frac{\alpha_0}{2\kappa^2}=\frac{f_X^0-X_0 f_{XX}^0}{2\kappa^2}\geq 0\;.$$

\end{itemize}  Recall that, for an $f(X)$-theory, the last three requirements are not independent.

\subsection{Equivalence Of $f(X,Y,Z)$ Gravity With Scalar-Tensor Theory}

It has been shown in Ref. \cite{chiba}, that the gravity theory described by $S\propto\int d^4x\sqrt{|g|}f(X,Y,Z)$ is equivalent to a multi-scalar-tensor gravity with the scalar fields coupled to curvature invariants ($R$, $C^2$, and ${\cal G}$). The equivalent action, however, is still very complicated and not useful to study the particle spectrum of the theory \cite{chiba}. An alternative is to start with the linearized action (\ref{lineaction}). Actually, by introducing an auxiliary field $\vphi$, the latter action can be rewritten in the following form \cite{chiba}: 

\be S\propto\int d^4x\sqrt{|\bar g|}\{\bar R-\frac{3}{2}(\bar\nabla\vphi)^2-V(\vphi)-\frac{1}{2m_2^2}\bar C^2\}\;,\label{chibaeq}\ee where $$\vphi=\ln(1+\frac{R}{3m_0^2})\;,$$ and the following scalar field self-interaction potential has been introduced:

\be V(\vphi)=\frac{3m_0^2}{2}(1-e^{-\vphi})^2+2\Lambda e^{-2\vphi}\;.\label{potential}\ee The metric $g_{\mu\nu}$ has been rescaled as $\bar g_{\mu\nu}=e^\vphi g_{\mu\nu}$ ($(\bar\nabla\vphi)^2\equiv \bar g^{\mu\nu}\nabla_\mu\vphi\nabla_\nu\vphi$). In what follows we consider that the ghost propagating mode associated with the Weyl squared tensor -- last term in the above action -- decouples from the gravitational spectrum. Otherwise, we assume $m_2^{-2}=0\;\Rightarrow\;f_Y^0=-4f_Z^0$ (see the definition of $m_2^2$ in equation (\ref{masses})). The latter requirement is automatically satisfied if the invariants $Y$ and $Z$ enter in the function $f(X,Y,Z)$ in the following combination: $Z-4Y$, as, for instance, in the Gauss-Bonnet invariant ${\cal G}$ \cite{comelli} (see also \cite{navarro}). As long as we will be concerned here with the mass of the scalar mode, this simplification will make easier further mathematical handling and physical interpretation of the results.

Looking at the potential (\ref{potential}) -- where the cosmological constant has a non-vanishing contribution -- and, since we started here with a linearization of the original theory and not with the theory itself, it might seem that the above equivalence among $\int d^4x\sqrt{|g|}f(X,Y,Z)$ and (\ref{chibaeq}) is not as strict as the similar equivalence between a given $f(R)$ theory and a dual scalar-tensor theory. At least one expects that the mass of the scalar field can be modified by the would be ghost excitation, even if it is decoupled from the gravitational spectrum \cite{ovrut}. However, as we will immediately show, the equivalence is strict. Our argument will rest on the computation of the effective mass of the scalar field. For this purpose let us to rewrite (\ref{lineaction}) as an $f(R)\equiv f(X)$-theory (recall that we have set $m_2^{-2}=0$):

\be S\propto\int d^4x\sqrt{|g|}\bar f(X),\;\;\bar f(X)=X+\frac{1}{6m_0^2}X^2-2\Lambda.\label{fr}\ee 

It has been shown that, in the weak field limit, looking for small spherically symmetric perturbations around de Sitter space with constant curvature $X_0=R_0$, the effective mass of the scalar degree of freedom $\phi\equiv\bar f_X$ in the equivalent scalar-tensor theory, is computed though the following equation (see, for instance, Ref. \cite{faraonirev}): $$m_{eff}^2=\frac{\bar f_X^0-X_0\bar f_{XX}^0}{3\bar f_{XX}^0}\;.$$ Hence, since for $\bar f(X)$ in (\ref{fr}); $$\bar f_X^0=1+\frac{X_0}{3m_0^2},\;\;\bar f_{XX}^0=\frac{1}{3m_0^2}\;,$$ then the effective mass of the scalar degree of freedom coincides with the mass of the spin-0 excitation $m_0^2$ computed through (\ref{masses}): $m_{eff}^2=m_0^2$. 

The above demonstration has several flags. Amongst them are the following: i) it rests on the weak field limit of the theory, and on a series expansion around maximally symmetric spaces of constant curvature, ii) it is based on a particular choice $m_2^{-2}=0$. The former drawback is common to standard demonstrations of equivalence between a given $f(R)$-theory and its scalar-tensor dual, while the latter one is not really a flag, but a necessary condition for the absence of Weyl ghosts in general $f(X,Y,Z)$-theories.

\subsection{Critical Comments}

Here we will take a step aside, to discuss the importance of a careful and consistent investigation of the picture generated by the expansion around maximally symmetric spaces of constant curvature discussed above in this section, and the consequent calculation of the masses of the excitations associated with the linearized theory.

At this point the first topic we want to comment on is the one related with the lack of coincidence in the expressions one finds in the bibliography for the masses (squared) of the additional propagating spin-2 and spin-0 degrees of freedom $m_2^2$ and $m_0^2$ (the same applies to the cosmological constant $\Lambda$ emerging after linearizing (\ref{action})). In Ref. \cite{comelli}, for instance, in the definition of the inverse mass squared coefficient $m_0^{-2}$, there appear terms $\propto R^2$ so that, when substituted in (\ref{linearized}) these give contributions of the fourth order in $R$, while the remaining coefficients $m_2^{-2}$ and $\Lambda$ produce quadratic contributions only. This points to an inconsistency of the order of the expansion around background spaces of constant curvature $R_0$ considered therein. Notice that if the terms $\propto R^2$ in the expression for $m_0^{-2}$ in Ref. \cite{comelli} are removed, the results are consistent with the ones reported above.

Another example is supplied by Ref. \cite{navarro}. In that reference the authors show expressions for the coefficients $\Lambda$ and $m_0^{-2}$ that are consistent with an expansion keeping all orders up to $R^4$, while $m_2^{-2}$ is consistent with keeping only up to $R^2$. The crucial point in this case is that, if in the expansion (\ref{taylor},\ref{taylor explicit}) one keeps all orders in $R$, i. e., up to terms $\sim R^4$ -- which is legitimate if one considers $R\sim 1$ -- then, not only the masses $m_0^2$ and $m_2^2$ will receive additional contributions but, at the same time, there will necessarily appear in (\ref{linearized}) additional terms proportional to $R^3$ and $R^4$ of the following form:

\bea \frac{a_{13}}{2}C^2R-\frac{a_{23}}{2}{\cal G}R+\frac{a_{33}}{3}R^3+\frac{a_{14}}{8}C^4-\frac{a_{24}}{4}C^2{\cal G}\nonumber\\
+\frac{a_{34}}{6}C^2R^2+\frac{a_{44}}{8}{\cal G}^2-\frac{a_{54}}{6}{\cal G}R^2+\frac{a_{64}}{18}R^4\;,\label{r4}\eea where the coefficients are given by:

\bea &&a_{13}=f_{YY}^0+4f_{ZX}^0,\;a_{23}=f_{YX}^0+2f_{ZX}^0,\nonumber\\
&&a_{33}=f_{YX}^0+f_{ZX}^0,\;a_{14}=f_{YY}^0+8f_{ZY}^0+16f_{ZZ}^0,\nonumber\\
&&a_{24}=f_{YY}^0+6f_{ZY}^0+8f_{ZZ}^0,\;a_{34}=f_{YY}^0+5f_{ZY}^0+4f_{ZZ}^0,\nonumber\\
&&a_{44}=f_{YY}^0+4f_{ZY}^0+4f_{ZZ}^0,\;a_{54}=f_{YY}^0+3f_{ZY}^0+2f_{ZZ}^0\nonumber\\
&&a_{64}=f_{YY}^0+2f_{ZY}^0+f_{ZZ}^0\;.\label{coeff}\eea 

These additional contributions towards (\ref{linearized}) are problematic and, as long as we know, their inclusion in the linearized action has not been investigated in detail yet, so that we do not really know how to deal with them. Notice, in between, that some expressions in \cite{comelli} and in \cite{navarro} coincide, while others do not.

To conclude these critical comments we want to underline that, in general, it is misleading keeping terms with order higher than $R^2$ in Taylor expansion, since, in particular, it arises an illusion that one can extrapolate the obtained formulas to get to qualitative conclusions about the behavior of the masses of the excitations at large $R$-s.  As we have shown, this is a wrong procedure since the definition of the masses of the excitations is given following the study of the linearized action \cite{stelle,ovrut} 

\be S\propto\int d^4x\sqrt{|g|}\left(-2\Lambda+R+\frac{1}{6m_0^2}R^2-\frac{1}{2m_2^2}C^2\right),\label{lineaction}\ee while consideration of terms $\propto R^3, R^4$ yields that the above linearized action is complemented with the addition of the terms in (\ref{r4}). The actual situation is in fact a bit more dramatic; at large $R$-s the linearized action above is replaced by one of the following form:
   
\bea &&S\propto\int d^4x\sqrt{|g|}\{R^4+\alpha(C^2+\beta {\cal G})R^2\nonumber\\&&\;\;\;\;\;\;\;\;\;\;\;\;\;\;\;\;\;\;\;\;\;\;\;\;\;\;\;\;+\gamma C^4+\delta C^2 {\cal G}+\mu{\cal G}^2\},\nonumber\eea where $\alpha$, $\beta$, $\gamma$, $\delta$, and $\mu$ are constant parameters. The latter action shares nothing in common with the linearized action above.

\section{Dirac-Born-Infeld Lagrangians}

In this section we shall explore the consequences of applying the DBI strategy to $f(X,Y,Z)$ theories, from the particular viewpoint of their stability properties. In this regard it is of prime importance to study the perturbations of the theory about background spaces of constant curvature. In fact it suffices to consider expanding the action

\be S_{DBI}\propto\int d^4x\sqrt{|g|}\lambda\left(\sqrt{1+\frac{2f(X,Y,Z)}{\lambda}}-1\right),\label{dbi action}\ee about maximally symmetric spaces of constant curvature $X=X_0=R_0$, up to $\sim (X^i-X_0^i)^2$, and keeping up to $\sim X^2\approx X_0^2$. As discussed in section II, the latter is an independent requirement having nothing to do with the order of the expansion around $X_0,Y_0,Z_0$. One obtains the following expression to determine the masses of the propagating spin-0 and spin-2 excitations (compare with Eq. (\ref{masses})):

\bea &&m_0^2=\frac{(\lambda+2f_0)(f_X^0-X_0f_{XX}^0)+X_0(f_X^0)^2}{(\lambda+2f_0)(3f_{XX}^0+2f_Y^0+2f_Z^0)-3(f_X^0)^2},\nonumber\\
&&m_2^2=-\frac{(\lambda+2f_0)(f_X^0-X_0f_{XX}^0)+X_0(f_X^0)^2}{(\lambda+2f_0)(f_Y^0+4f_Z^0)}\;.\label{dbi masses}\eea In the formal limit $\lambda\rightarrow\infty$ -- unbounded curvature -- one recovers the known expressions for $m_0^2$ and $m_2^2$ in Eq. (\ref{masses}). 

Notice that, independent on the scale $\lambda$, the necessary condition to avoid the Weyl ghost is the same as before: $f_Y^0=-4f_Z^0$, so that the DBI modification plays no role in avoiding the occurrence of massive, spin-2 ghosts modes arising in non-linear theories of the kind $f(X,Y,Z)$. 

The expressions for the masses of the propagating linear modes are important since the stability of these modes depends crucially on their sign. Hence, in the first place, the DBI modification affects the stability properties of the gravitational spectrum of the propagating excitations. Under the DBI modification of the original $f(X,Y,Z)$-theory, the trace equation (\ref{trace'}) is replaced by the following:

\be 2\lambda\left(1+\frac{2f_0}{\lambda}-\sqrt{1+\frac{2f_0}{\lambda}}\right)-X_0f_X^0-2Y_0f_Y^0-2Z_0f_Z^0=0.\label{dbi trace}\ee Once again, for unbounded curvature $\lambda\rightarrow\infty$, equation (\ref{trace'}) is recovered. The above trace equation is usually an algebraic equation to determine the curvature $X_0$ of the maximally symmetric background space about which we are perturbing the original equations of the theory.

In the next subsections A and B, we investigate a couple of examples taken from the literature to illustrate the discussion above. In fact we shall examine only the absence of tachyon instability and of the Weyl ghost, which are just an aspect of the stability of these theories. Then, in subsection C, we discuss in more detail other relevant aspects of the stability issue, and apply the results to these examples.

\subsection{Example I}

Take as an example the $f(X,Y,Z)$ theory studied in \cite{navarro} (proposed in \cite{carroll} in a cosmological setting):

\be f(X,Y,Z)=X-\frac{\mu^{4n+2}}{(a X^2+b Y+c Z)^n}\;,\label{acoleyen}\ee where $\mu$ is a parameter with dimensions of mass, and $n$ is positive. In the present example the trace condition (\ref{trace}) for a maximally symmetric space of constant curvature $X=X_0$, amounts to 

\be X_0^{2n+1}=\frac{2(n+1)\mu^{4n+2}}{k^n}\;,\label{x0}\ee where we have introduced the parameter $k\equiv a+b/4+c/6$. Hence, considering large or small curvatures is not at will. Actually, given a fixed set of overall parameters $a$, $b$, $c$, $n$, and $\mu$, the value of the curvature $R_0$ is fixed through (\ref{x0}), which is the unique root of the trace equation (\ref{trace'}) for $f(X,Y,Z)$ given by (\ref{acoleyen}). In this case, for the masses of the propagating excitations one obtains, according to (\ref{masses}): 

\bea &&m_0^2=p\left(\frac{n+1}{nl}\right)(k^2+2na^2)\mu^2\;,\nonumber\\
&&m_2^2=p\left(\frac{n+1}{nk}\right)\left(\frac{k^2+2na^2}{b+4c}\right)\mu^2\;,\label{acoleyen masses}\eea where $$l\equiv (3a+b+c)k-6(n+1)a^2,$$ and $$p\equiv\left[\frac{2(n+1)}{k^n}\right]^{1/(2n+1)}.$$ It is curiously enough that, in the present case, for the corresponding DBI modified theory (\ref{dbi action}), the expansion around spaces of constant curvature $X_0$ yields the same results for the masses of the spin-0 and spin-2 excitations as in Eq. (\ref{acoleyen masses}).\footnote{To be consistent with the order of the approximation, one has to neglect terms $\propto X_0^3\sim X_0^2\lambda^{-1}$ and higher.} This means that, for the theory (\ref{acoleyen}), up the given approximation, the stability of the perturbations is not affected by the DBI modification (\ref{dbi action}).

\subsection{Example II}

To illustrate how the DBI modification does evidently affect the dynamics, we choose a theory that fulfills the requirements to be phenomenologically viable, listed at the beginning of the section. It is based on the following DBI-type action, which represents a minimal deformation (a trivial change of signs only) of the one in \cite{comelli}:

\bea &&S\propto\int d^4x\sqrt{|g|}\frac{1}{\kappa^2}\left(\sqrt{1+2\kappa^2 f(X,Y,Z)}-1\right),\nonumber\\
&&\;\;\;\;\;\;\;\;\;\;f(X,Y,Z)=\alpha X-\kappa^2\beta (X^2-4Y+Z).\label{action comelli}\eea where $\lambda=\kappa^{-2}=M_{Pl}^2$ is the maximum curvature scale, while $\alpha$, and $\beta$ are overall constants. Notice that the ``unmodified" action $S\propto\int d^4x\sqrt{|g|}f(X,Y,Z)$ coincides with the Einstein-Hilbert one, since the Gauss-Bonnet term $X^2-4Y+Z$ contributes a total divergence that can be safely erased from the action. Hence, in the linearized approximation, there is only one massless spin-2 propagating excitation (properly the graviton), while the gravitational spectrum of the theory depicted by the action (\ref{action comelli}), in the linearized limit, consists of a massless spin-2 propagating degree of freedom, and of a spin-0 (massive) excitation. Otherwise, it is properly a scalar-tensor theory of gravity. In this case the DBI procedure evidently affects the ``unmodified" $f(X,Y,Z)$ theory, even at the linearized level. The mass of the scalar degree of freedom in the linear approximation is given by (see Eq. (\ref{dbi masses})):

\be m_0^2=\frac{1+3\alpha\kappa^2X_0}{3\kappa^2(4\beta\kappa^2X_0-\alpha)},\label{dbi mass}\ee where we have dropped terms of order $\propto X_0^3,\kappa^2X_0$ and higher. The value of the background curvature is a real root of the trace equation (\ref{dbi trace}). Notice that in the unbounded curvature limit $\kappa^2\rightarrow 0$ ($\lambda\rightarrow\infty$), the mass of the scalar perturbation blows up and is negative, so that this limit, if it exist in the present theory, is largely unstable. Besides, in order for the theory (\ref{action comelli}) to be free of tachyon instability (negative $m^2$-s), unless $\alpha$ is negative, the value of the background curvature is bounded from below $$X_0\geq \frac{\alpha}{4\beta\kappa^2}\;.$$ On the other hand, the constant $\alpha$ can not be negative because, in the low curvature limit, it modifies the sign of the the effective Newton's constant $G=(8\pi)^{-1}M_{Pl}^{-2}$. This demonstrates that the theory given by (\ref{action comelli}) does admit stable perturbations around maximally symmetric spaces of constant curvature (otherwise, it does admit stable de Sitter solutions), only for unnaturally small values of the constant $\alpha$: $\alpha\sim\beta\kappa^2H_0^2=\beta M_{Pl}^{-2}H_0^2$ ($H_0^2$ is the current value of the Hubble parameter), so that we are faced with a very serious problem of fine tunning. Although possible, this is a very unlikely scenario. Of course, the situation here is that the stability of the theory is very sensitive to the signs in (\ref{dbi modification},\ref{dbi action}). If one chooses an alternative deformation {\it a la} Dirac-Born-Infeld as it follows (compare with (\ref{dbi modification}) or, equivalently, with (\ref{dbi action}) and note the change of signs):

\be f(X,Y,Z)\rightarrow\lambda\left(1-\sqrt{1-2f(X,Y,Z)/\lambda}\right)\;\label{dbi alternative}\ee then, after expanding the corresponding action around a maximally symmetric space of constant curvature $X=X_0$, one is led to the following expression to determine the mass of the spin-0 excitation (compare with (\ref{dbi masses})): 

\be m_0^2=\frac{(\lambda-2f_0)(f_X^0-X_0f_{XX}^0)-X_0(f_X^0)^2}{(\lambda-2f_0)(3f_{XX}^0+2f_Y^0+2f_Z^0)+3(f_X^0)^2}\;.\label{dbi masses alternative}\ee In the theory with $f(X,Y,Z)$ given by (\ref{action comelli}), one obtains (compare with Eq. (\ref{dbi mass})):

\be m_0^2=\frac{1-3\alpha\kappa^2X_0}{3\kappa^2(\alpha-4\beta\kappa^2X_0)}\;,\label{dbi mass alternative}\ee where, as before, to be consistent with the approximation undertaken, we have dropped terms of order $\propto X_0^3,\kappa^2X_0^2,\kappa^4 X_0$ and higher. The tachyon instability is absent if either: i) $X_0<1/3\kappa^2\alpha$ and, at the same time, $X_0<\alpha/4\kappa^2\beta$, or ii) $X_0>1/3\kappa^2\alpha$ and, at the same time, $X_0>\alpha/4\kappa^2\beta$. The latter possibility is very unlikely to occur since, to obtain a stable scalar perturbation, the constant $\alpha$ has to be unnaturally small $\alpha\sim\beta M_{Pl}^{-2}H_0^2$ (see similar discussion above). The former possibility is indeed appropriate since it is compatible with a maximum curvature, as it is expected for DBI models. For $\alpha^2>\beta$, the limiting (larger available) curvature is $X_0^{max}=1/3\kappa^2\alpha$, while for $\beta>\alpha^2$ it is $X_0^{max}=\alpha/4\kappa^2\beta$. So, perhaps, the alternative DBI strategy (\ref{dbi alternative}) is more attractive than the first one (\ref{dbi modification}) in the present case.

\subsection{DBI Modification of Stability}

Here, we will summarize the way the DBI strategy modifies the stability requirements listed in subsection A of section II for arbitrary functions $f(X,Y,Z)$. For this purpose we write the two possible choices of the DBI modification in the following general form:

\be f\rightarrow\epsilon\lambda(\sqrt{1+2\epsilon f/\lambda}-1)\;,\label{dbi united}\ee where $\epsilon=\pm 1$ ($\epsilon^2=1$). Written in this form, equation (\ref{dbi united}) comprises both: procedure given by Eq. (\ref{dbi modification}) if one chooses $\epsilon=+1$, and the alternative procedure given by Eq. (\ref{dbi alternative}) if one chooses $\epsilon=-1$. The following relationships are useful to study the considered modifications of the stability:

\bea &&f_X\rightarrow\Gamma f_X\;,\;\;f_Y\rightarrow\Gamma f_Y\;,\;\;f_Z\rightarrow\Gamma f_Z\;,\nonumber\\
&&f_{XX}\rightarrow\Gamma\left(f_{XX}-\frac{\epsilon}{\lambda}\Gamma^2f_X^2\right)\;,\nonumber\eea where the DBI ``boost'' is defined as $$\Gamma\equiv\Gamma(X,Y,Z)=\left(1+\frac{2\epsilon}{\lambda}f\right)^{-1/2}\;.$$ Hence, after applying the DBI deformation to quadratic modifications of Einstein-Hilbert theory, the following modifications of the stability requirements occur (compare with similar requirements in subsection A of section II):

\begin{itemize}

\item{Ostrogradski Stability:} The same as before, i. e., it is unaffected by the DBI procedure $$f_Y^0+4f_Z^0=0\;.$$

\item{Ricci Stability:} $$f_{XX}^0-\frac{\epsilon}{\lambda}\Gamma_0^2 (f_X^0)^2\geq 0\;.$$ 

\item{Absence of Tachyon Instability:} $$m_0^2=\frac{f_X^0-X_0 f_{XX}^0+\frac{\epsilon}{\lambda}X_0\Gamma_0^2 (f_X^0)^2}{3f_{XX}^0+2f_Y^0+2f_Z^0-\frac{3\epsilon}{\lambda}\Gamma_0^2(f_X^0)^2}\geq 0\;.$$

\item{Non-negativity of the Gravitational Coupling:} $$f_X^0-X_0 f_{XX}^0+\frac{\epsilon}{\lambda}X_0\Gamma_0^2 (f_X^0)^2\geq 0\;.$$

\end{itemize} Additionally it is required that the DBI boost be a real quantity, i. e., $$\Gamma\in\Re\;\;\Rightarrow\;\;2\epsilon f+\lambda\geq 0\;.$$ Notice that, in the formal limit $\lambda\rightarrow\infty$, the above requirements on the stability coincide with the ones in section II A.

Let us to check the theory (\ref{acoleyen}) \cite{navarro}, regarding other, perhaps more restrictive, stability criteria. For instance, since in order to avoid propagating Weyl ghosts $b=-4c\;\Rightarrow\;k=a-5c/6$, then $$f_{XX}^0=-\frac{2na[(2n+1)a+5c/6]\mu^{4n+2}}{k^{n+2}X_0^{2n+2}}<0\;,$$ therefore, for the space of parameters allowed by the observations, the theory of Ref. \cite{navarro} is Ricci unstable, so that it develops ghosts. This argument might be enough to rule out this theory. Next see what happens if we modify it {\it a la} Dirac-Born-Infeld. Would it be yet Ricci unstable? After applying the DBI deformation procedure (\ref{dbi united}), the above condition on $f_{XX}$ translates into the following bound: $$X_0^{2n}[\lambda\bar k+\epsilon a(n+1)X_0]\leq\frac{\epsilon[6(n+1)a+5c]}{3k^n}\mu^{4n+2}\;,$$ where $\bar k\equiv(2n+1)a+5c/6$. For small $X_0\ll\lambda$, we have $$X_0^{2n}\leq\frac{\epsilon[6(n+1)a+5c]}{3\bar k k^n\lambda}\mu^{4n+2}\;,$$ which is consistent only for DBI embedding with $\epsilon=+1$. Notice that in the formal limit $\lambda\rightarrow\infty$ the above bound can not be satisfied and we get again a Ricci unstable theory. For the opposite embedding, i. e., $\epsilon=-1$, the only way to achieve Ricci stability, is that $X_0>\lambda\bar k/(n+1)a$, leading to the following constraint: $$X_0^{2n}[(n+1)a X_0-\lambda\bar k]\geq\frac{6(n+1)+5c}{3k^n}\mu^{4n+2}\;.$$ In this case, $\lambda$ is not an upper curvature bound any more, and the original motivation of the DBI strategy is lost.

We see that the DBI modification strategy indeed modifies the Ricci stability criterion as well, making a originally Ricci unstable theory, a stable one.

\section{DBI Deformation Strategy Applied to $f(R)$ Theories}

We have seen how the application of the DBI procedure -- both
variants: (\ref{dbi modification}), and (\ref{dbi alternative}) --
actually modifies the stability properties of $f(X,Y,Z)$ gravity
theories\footnote{This is the usual case in DBI modifications,
  see\cite{Copeland:2010jt} for an example of a DBI-modified cosmological scalar field.}. Even the gravitational spectrum of the theory is altered if one follows this strategy with a class of theories given by $f(X,Y,Z)=X+\alpha{\cal G}$ (these include the particular case $\alpha=0$). Here we will be concerned with possible modifications of the stability of de Sitter backgrounds in the simpler $f(X)$(or $f(R)$) gravity. In what follows, just for homogeneity of writing, we keep the variable $X$ instead of $R$, although both will be used interchangeably. Besides, for generality of the discussion we will consider the DBI deformation procedure given in Eq. (\ref{dbi united}). The field equations that can be obtained from an action principle, the action being:

\be S=\frac{1}{2\kappa^2}\int d^4x\sqrt{|g|}f(X)+S_{(m)}(g_{\mu\nu},\psi)\;,\label{fr action}\ee where $S_{(m)}$ is the action of matter, and $\psi$ - the collective name for the matter degrees of freedom, can be written in the following form:

\be f'(X)\;G_{\mu\nu}=\kappa^2\left(T_{\mu\nu}^{(m)}+T_{\mu\nu}^{(cur)}\right)\;,\label{fr feqs}\ee where the comma denotes derivative with respect to the curvature $X$, and we have introduced the following definition of the effective curvature tensor:

\bea &&\kappa^2 T_{\mu\nu}^{(cur)}\equiv\frac{f(X)-Xf'(X)}{2}\;g_{\mu\nu}\nonumber\\
&&\;\;\;\;\;\;\;\;\;\;\;\;\;\;\;\;\;\;\;\;\;+\nabla_\mu\nabla_\nu f'(X)-g_{\mu\nu}\Box f'(X)\;.\label{eff stress}\eea The trace of Eq. (\ref{fr feqs}) amounts to an additional constraint on $X$:

\be Xf'(X)-2f(X)+3\Box f'(X)=0\;.\label{fr trace'}\ee If one expands the action (\ref{fr action}) -- considering vacuum background, i. e., $S_{(m)}=0$ -- around maximally symmetric spaces of constant curvature $X=X_0$, just as it has been done in section III, then one obtains:

\be S\propto\alpha_0\int d^4x\sqrt{|g|}\left(-2\Lambda+X+\frac{1}{6m_0^2}X^2\right)\;,\label{fr linearized}\ee where ($f_0\equiv f(X_0)$, etc.) $\alpha_0\equiv f'_0-X_0 f''_0$, and

\bea &&-2\Lambda\equiv\frac{f_0-X_0f'_0+X_0^2 f''_0/2}{f'_0-X_0 f''_0},\nonumber\\
&&\;\;\;m_0^2\equiv\frac{f'_0-X_0 f''_0}{3 f''_0}\;.\label{fr mass}\eea The gravitational spectrum consists of a massless spin-2 graviton, and a massive spin-0 propagating degree of freedom, with mass squared $m_0^2$. In this case, the trace equation (\ref{fr trace'}) translates into an algebraic equation to determine $X_0$: 

\be X_0 f'_0=2f_0\;.\label{fr trace}\ee As it has been already explained, requirements of Ricci stability, absence of tachyon instability, and non-negativity of the effective gravitational coupling are not independent requirements in the present case, so that one has to check only two of them, for instance, Ricci stability, and non-negativity of the gravitational coupling:

\be f''\geq 0,\;\;f'_0-X_0 f''_0\geq 0\;.\label{stability criteria}\ee

The same linearization around vacuum backgrounds with constant curvature, when applied to $f(X)$ gravity which has suffered further DBI deformation given by Eq. (\ref{dbi united}), yields to the following expression for the mass of the spin-0 propagating excitation (compare with Eq. (\ref{fr mass})):

\be m_0^2=\frac{f'_0-X_0f''_0+\frac{\epsilon}{\lambda}\Gamma_0^2 X_0 (f'_0)^2}{3f''_0-\frac{3\epsilon}{\lambda}\Gamma_0^2 (f'_0)^2}\;,\label{dbi fr mass}\ee where $$\Gamma_0=\frac{1}{\sqrt{1+2\epsilon f_0/\lambda}}\;,$$ while the trace equation (\ref{fr trace}) translates into the following equation:

\be \Gamma_0^2 X_0 f'_0=2\epsilon\lambda(1-\Gamma_0)\;.\label{dbi fr trace}\ee Therefore, the stability criteria (\ref{stability criteria}) translate into the following requirements:

\bea &&f''_0-\frac{\epsilon}{\lambda}\Gamma_0^2 (f'_0)^2\geq 0\;,\nonumber\\
&&f'_0-X_0f''_0+\frac{\epsilon}{\lambda}\Gamma_0^2 X_0 (f'_0)^2\geq 0\;.\label{dbi stability criteria}\eea

In what follows we will explore the consequences for stability of -- constant curvature -- vacuum background linearization, of applying the DBI procedure given by Eq. (\ref{dbi united}) to $f(X)$ theories of gravity. We plan do that by working out two examples taken from the bibliography.

\subsection{Example I}

We start the study of the possible modification of the stability of (constant curvature) vacuum background linearization of $f(X)$ theories -- which have been further modified according to the DBI deformation strategy --, by choosing the model of reference \cite{positive}, that is given by:

\be f(X)=X-\frac{\mu^{2(n+1)}}{X^n}\;,\label{fr1}\ee where $\mu$ is a suitably chosen parameter. A check of the stability criteria (\ref{stability criteria}) shows that $$f''_0=-\frac{n(n+1)\mu^{2(n+1)}}{X_0^{n+2}}\;,$$ is negative, which means that the model suffers from Ricci scalar instability \cite{kawasaki,faraoni2006,faraoni2007} (see also \cite{faraonirev,nojiri}). Nonetheless, $$f'_0-X_0f''_0=1+\frac{n(n+2)\mu^{2(n+1)}}{X_0^{n+1}}\geq 0\;,$$ so that positivity of the effective gravitational coupling is guaranteed. As a consequence (see the definition of the spin-0 mode mass squared in Eq. (\ref{fr mass})), the model suffers also from the scalar tachyon instability. As explained before, the Ricci instability is enough to rule out the model since this type of instability develops very quickly.

What one should expect from applying the DBI strategy in this case? The answer can be based on the analysis of the DBI modification of the stability criteria in Eq. (\ref{dbi stability criteria}). According to these, the Ricci stability bound is given, in the present case, by the following constraint: $$-\frac{X_0^{n+1}}{n(n+1)\mu^{2(n+1)}}-\frac{\epsilon\lambda}{X_0}+\left(\frac{n+2}{n+1}\right)\left(\frac{\mu^{2(n+1)}}{X_0^{n+1}}-2\right)\geq 0.$$ For small curvature $X_0^{n+2}\ll\lambda\mu^{2(n+1)}$, the above constraint simplifies $$-\frac{\epsilon\lambda}{X_0}+\left(\frac{n+2}{n+1}\right)\left(\frac{\mu^{2(n+1)}}{X_0^{n+1}}-2\right)\geq 0\;,$$ so that for the DBI embedding with $\epsilon=-1$, whenever $X_0^{n+1}<\mu^{2(n+1)}/2$, this bound is always satisfied. Besides, if in order to simplify the analysis, one assumes that $X_0\ll\mu^2$, then, the criterion requiring non-negativity of the effective gravitational coupling upon linearization, can be written in the form of the following bound: $$\lambda-\epsilon\left(\frac{n+4}{n+2}\right)\frac{\mu^{2(n+1)}}{X_0^{n+1}}\geq 0\;,$$ which for the $\epsilon=-1$ embedding is always satisfied. Therefore, the DBI deformed theory is not only Ricci stable, but also, it is free of scalar tachyon instability.

\subsection{Example II}

Our next example also reflects the consequences of applying the DBI deformation procedure (\ref{dbi united}). Let us consider the following $f(X)$ gravity theory \cite{amendola}:

\be f(X)=X-(1-n)\mu^{2(1-n)}X^n\;,\label{fr2}\ee where, to be compatible with observations $0<n\leq 0.25$, and $\mu$ is a sufficiently small parameter ($\mu\sim 10^{-50} eV$) \cite{faulkner}. If one linearizes the corresponding action $S\propto\int d^4x\sqrt{|g|}f(X)$, then one finds that the mass of the spin-0 excitation Eq. (\ref{fr mass}), in the present case is given by:

\be m_0^2=-\left(\frac{n-2}{n-1}\right)\left\{1-\frac{(X_0 \mu^{-2})^{1-n}}{n(n-1)(n-2)}\right\}\frac{X_0}{3}\;,\label{fr mass2}\ee so that, for positive $n$-s within the range where the model is compatible with the observations (see above), a tachyon instability develops ($m_0^2<0$). Notwithstanding, since in this case, for the mentioned range of parameters compatible with observations $$f''_0=n(n-1)^2\mu^{2(1-n)}X_0^{n-2}>0\;,$$ then the model is Ricci stable. One might think that a large enough time to develop the tachyon instability could make the theory compatible with the cosmological dynamics. However, the very tricky situation with this model, comes, precisely, from the fact that Ricci stability, absence of tachyon instability and non-negativity of the effectve gravitational coupling are not independent requirements. In this particular case (Ricci stable, but with tachyon instability), the effective gravitational coupling is negative, which is catastrophic for this model, and it has to be ruled out.

Would replacement of $f(X)$ by $\epsilon\lambda(\sqrt{1+2\epsilon f(X)/\lambda}-1)$ save the model? Let us start by checking Ricci stability. In this case, since $\mu^2$ is small enough, one might consider two important limiting situations to simplify the analysis. Assume first that $X_0\gg\mu^2$, so that the Ricci stability bound in (\ref{dbi stability criteria}) can be written as $$X_0^{2-n}\leq\epsilon n(1-n)^2\lambda\mu^{2(1-n)}\;,$$ which is satisfied only for the DBI embedding with $\epsilon=+1$. Now assume a different limit, $X_0\sim\mu^2$. In this case, the above bound may be written in the form of the following constraint: $$\epsilon n(1-n)^2\lambda-p X_0\geq 0\;\Rightarrow\;X_0\leq\frac{\epsilon n(1-n)^2}{p}\lambda\;,$$ where the constant $p\equiv 1+n(1-n)^2-2n(1-n)(2-n)$ is positive for the range of parameters allowed by the observations. This means, once again, that the only DBI embedding that is enable to save the Ricci stability is the one with $\epsilon=+1$. The next step is to check for non-negativity of the effective gravitational coupling. In this case, for the limiting situation when $X_0\gg\mu^2$, the criterion of non-negativity of the effective gravitational coupling in (\ref{dbi stability criteria}) amounts to the following inequality: $$X_0^n\leq\frac{\epsilon\lambda}{2(1-n)(1+3n-n^2)}\;,$$ which is obeyed whenever $\epsilon=+1$ as it should be according to the previous results on Ricci stability. If one considers, instead, the limiting situation when $X_0\sim\mu^2$, one obtains that the effective gravitational coupling is non-negative if $$X_0\leq\epsilon\lambda\frac{k}{l}\;,$$ where the constants $k\equiv 1-n(1-n)(2-n)$, and $l\equiv 2(1-n)(1+3n-n^2)-n(1-n)^2(4-n)$, are positive for the range of parameters allowed by the observations. Hence, once again, the bound is satisfied if $\epsilon=+1$.

We have seen that, in general, the DBI strategy to modify $f(X)$ theories can help avoiding either ghosts or the tachyon instability or, it can even help avoiding both simultaneously. For Lagrangians where the quadratic contributions come with inverse powers of the curvature, the correct DBI embedding is the one with $\epsilon=-1$, while for quadratic contributions proportional to positive powers of the curvature, the correct embedding is the one with $\epsilon=+1$. In the next section we will study the modifications to the cosmic dynamics caused by applying the DBI strategy to $f(X,...)/f(X)$-theories.

\section{Cosmological Consequences of Applying the DBI Strategy}

In this section we will explore the impact of applying the DBI procedure -- both possibilities: (\ref{dbi modification},\ref{dbi action}), and (\ref{dbi alternative}) -- on the cosmological dynamics of $f(X,...)$/$f(X)$ gravity theories. Due to their simplicity, we focus the discussion in $f(R)$ gravity theories exclusively. For the study of the impact of the DBI strategy applied to $f(X,...)$ theories, we refer the reader to the paper \cite{quiros}, where this issue has been investigated in detail, for the case when the above deformation procedure is applied to theories of the kind $f(X,{\cal G})$. We choose the Friedmann-Robertson-Walker (FRW) metric as a local description of spacetime at cosmological scales (for simplicity, without loss of generality, we consider FRW spaces with spatial sections of constant curvature):\footnote{For a discussion of the subtleties associated with the validity of the Ehlers-Geren-Sachs theorem \cite{ehlers} on the identification of FRW spacetime with our universe, as a consequence of the high degree of isotropy of the cosmic microwave radiation, we refer the reader to \cite{faraonirev} and linked references therein.}

\be ds^2=-dt^2+a^2(t)\delta_{ij}dx^idx^j\;.\label{frw}\ee Inserting this choice of the metric into the field equation (\ref{fr feqs}), and assuming a perfect fluid stress-energy tensor $$T_{\mu\nu}^{(m)}=(\rho_m+p_m)u_\mu u_\nu+p_m g_{\mu\nu}\;,$$ where $u^\mu$ is the fourth-velocity of an observer co-moving with the fluid, $\rho_m$ and $p_m$ being the energy density and the pressure of the fluid, respectively, to describe the macroscopic behavior of matter, one obtains the following cosmological equations:

\bea &&3H^2=\frac{\kappa^2}{f'}(\rho_m+\rho_{cur})\;,\nonumber\\
&&2\dot H+3H^2=-\frac{\kappa^2}{f'}(p_m+p_{cur})\;,\label{frw feqs}\eea where the effective (parametric) energy density $\rho_{cur}$ and pressure $p_{cur}$ of the effective ``curvature" fluid, are given by:

\bea &&\rho_{cur}=\frac{1}{\kappa^2}\left[\frac{1}{2}(X f'-f)-3H\dot f'\right]\;,\nonumber\\
&&p_{cur}=\frac{1}{\kappa^2}\left[2H\dot f'+\ddot f'-\frac{1}{2}(X f'-f)\right]\;,\label{rho p}\eea respectively. As it is clear from going to the limit $\rho_m\rightarrow 0$ in the first equation in (\ref{frw feqs}) -- properly the Friedmann equation --, the effective energy density $\rho_{cur}$ can not be negative \cite{faraonirev}. It can be defined also an effective equation of state (EOS) parameter $w_{cur}\equiv p_{cur}/\rho_{cur}$ in the following way:

\be w_{cur}=\frac{2(\ddot f'-H\dot f')}{X f'-f-6H\dot f'}-1\;,\label{eos}\ee from which one may see that the vacuum value $w_{cur}=-1$ is attained whenever $$\ddot f'=H\dot f'\;.$$ de Sitter solutions trivially satisfy this equality, but there is a class of non-de Sitter solutions for which $\dot f'\propto a(t)$ which also satisfy the above equation. 

Under the replacement (\ref{dbi united}), $\ddot f'$, $\dot f'$, and $f'$, transform like:

\bea &&\ddot f'\rightarrow\Gamma\left\{\ddot f'-\frac{\epsilon\Gamma^3}{\lambda}f'^2\left(\ddot X+3\frac{\dot f'}{f'}\dot X-3\frac{\epsilon\Gamma^2}{\lambda}f'\dot X^2\right)\right\}\;,\nonumber\\
&&\dot f'\rightarrow\Gamma\left(\dot f'-\frac{\epsilon\Gamma^2}{\lambda}f'^2\dot X\right)\;,\;\;f'\rightarrow\Gamma f'\;,\label{transform}\eea respectively, where $\Gamma=(1+2\epsilon f/\lambda)^{-1/2}$. Hence, the effective energy density and pressure of the ``curvature fluid'' (Eq. (\ref{rho p})), will be replaced by very complicated expressions. For that reason, to simplify the analysis, we will operate in a different, yet equivalent, way: we keep equations (\ref{frw feqs}), (\ref{rho p}), (\ref{eos}), valid for some $\bar f_\epsilon(X)$, such that $$\bar f_\epsilon(X)=\epsilon\lambda(\sqrt{1+2\epsilon f(X)/\lambda}-1)\;,$$ where $f(X)$ is the original, unchanged function, while $\bar f_\epsilon(X)$ will be the DBI modified one. Otherwise, to look for the impact of the DBI modification strategy on the cosmic dynamics of $f(X)$ models, one has just to replace $f(X)\rightarrow\bar f_\epsilon(X)$ in the mentioned equations. In the next subsection we will investigate the dynamics of given $f(X)$ models, to uncover the way the DBI procedure operates to modify it. In what follows, for simplicity, we will consider pressureless dust matter as the background fluid, i. e., $p_m=0$.

\subsection{Dynamical Systems Study}

Now, since finding exact solutions of the equations (\ref{frw feqs}) is in general a very difficult task, we will rely on the dynamical systems tools to investigate the asymptotic structure of the $f(X)$ models of interest, instead. To this end we will apply the concise recipes given in the appendix (section VIII). The goal will be to write the system of cosmological equations in the form of an autonomous system of ODE (as described in the appendix), so that one could associate such important dynamical systems concepts as past and future attractors (also saddle equilibrium points), with dynamical configurations -- solutions -- of the models. This is a powerful approach to uncover the most generic classes of solutions that are allowed by them. In order to build an autonomous system of ordinary differential equations (ODE) out of (\ref{frw feqs},\ref{rho p}), following \cite{amendola}, we introduce the following dimensionless variables:

\be x\equiv\frac{X}{6H^2}=\frac{\dot H}{H^2}+2\;,\;\;y\equiv-\frac{\dot f'}{H f'}\;.\label{psv}\ee In terms of these variables the Friedmann equation (first equation in (\ref{frw feqs})) can be written in the form of the following constraint:

\be \hat\Omega_m\equiv\frac{\kappa^2\rho_m}{3f'H^2}=1-x-y+\frac{f}{6H^2 f'}\;,\label{f c}\ee where we have conveniently defined an ``effective'' dimensionless matter energy density parameter $\hat\Omega_m$, which, thanks to the $f'$ function entering in its definition, might be, in principle, any sign, i. e., either positive or negative, without entering in conflict with standard physical requirements. Particular properties of a given model are encoded in the last term in the RHS of Eq. (\ref{f c}). In fact, not all of the variables of the phase space needed to describe the dynamics of a given model have been defined yet. This requires knowledge of the concrete model. Other general, useful expressions are 

\bea &&-\frac{\ddot f'}{H^2 f'}=2(x-2)+y+6\hat\Omega_m,\nonumber\\
&&w_{cur}=-1-\frac{2(x-2)+6\hat\Omega_m}{3\left[x-\frac{f/f'}{6H^2}+y\right]}.\label{useful}\eea The phase space variables $x$, $y$ obey the following autonomous system of ODE:

\bea &&\frac{dx}{d\tau}=\frac{dX/d\tau}{6H^2}-2x(x-2),\nonumber\\
&&\frac{dy}{d\tau}=(x-2)(2-y)+y+y^2+6\hat\Omega_m,\label{general ode}\eea where, as customary, we have introduced the time-ordering variable $\tau=\ln a$ -- basically the number of e-foldings. Depending on the concrete model, new variables have to be introduced and, as a consequence, the above equations have to be complemented with the addition of new ones (one per each new phase space variable). The particular properties of a given model enter in equations (\ref{general ode}) through the terms $dX/6H^2 d\tau$ and $\hat\Omega_m$ (see Eq. (\ref{f c})). Now we are in position to study a concrete model.

\subsection{Model $f(X)=\mu X^n$}

Although, in general ($n\neq 1$), it does not contain Einstein-Hilbert gravity as the low curvature limit (as it should be), the above model is simple enough, and has been formerly studied, for instance in \cite{amendola} (IJMPD). The related model $f(X)=X+\mu X^n$ -- also studied in \cite{amendola} (PRL) -- is a bit more realistic, yet much more complicated from the point of view of its asymptotic properties. However, the real complexity arises when one tries to study the dynamics of its DBI deformation. This will deserve a separate publication. That is the reason why we have not included this toy model as an example in this subsection. 

In the model $f(X)=\mu X^n$, $\mu$ and $n$ are the overall free parameters. In this case, since $f/f'=X/n$, hence $$\hat\Omega_m=1+\left(\frac{1-n}{n}\right)x-y\;.$$ Besides, since $$\dot f'=n(n-1)\mu X^{n-2}\dot X\;,$$ then $$\frac{dX}{d\tau}=\left(\frac{1}{1-n}\right)y X\;.$$ Using the above relationships, one can write (\ref{general ode}) in an explicit form:

\bea &&\frac{dx}{d\tau}=\left(\frac{1}{1-n}\right)xy-2x(x-2),\nonumber\\
&&\frac{dy}{d\tau}=6\left(1+\frac{1-n}{n}x\right)+(x-2)(2-y)\nonumber\\
&&\;\;\;\;\;\;\;\;\;\;\;\;\;\;\;\;\;\;\;\;\;\;\;\;\;\;\;\;\;\;\;\;\;\;\;\;\;\;\;\;\;\;\;\;\;\;\;\;\;\;\;\;-5y+y^2.\label{ode 1}\eea The above is a closed system of ODE, and one does not need more variables to describe the phase space dynamics, otherwise, the autonomous system (\ref{ode 1}) is defined within the following (open) 2-dimensional phase space:\footnote{Here we will consider only positive scalar curvature.} $$\Psi=\{(x,y): x\geq 0,\;\hat\Omega_m(x,y)\geq 0\}\;.$$ The latter bound is justified only if one assumes the parameters $\mu$ and $n$ to be both positive magnitudes, so that $f'$ is always non-negative. Other two magnitudes of immediate physical (and observational) meaning are, the deceleration parameter $q=1-x$, and the effective equation of state parameter of the ``curvature fluid'' $$w_{cur}=-1-\frac{2(x-2)+6\hat\Omega_m}{3\left(\frac{n-1}{n}\right)x+3y}\;.$$

The critical points $P_i=(\bar x_i,\bar y_i)$ of the autonomous system of ODE (\ref{ode 1}), together with their relevant properties, are listed below:

\begin{enumerate}

\item Curvature-dominated equilibrium point  $P_{rad}=(0,1)$, associated with decelerated peace of expansion. The ``curvature fluid'' mimics radiation. This critical point is characterized by the following values of the relevant parameters: $$\hat\Omega_m=0,\;\;w_{cur}=\frac{1}{3},\;\;q=1.$$ The eigenvalues of the corresponding linearization matrix are (see the appendix) $$\lambda_1=-1,\;\;\lambda_2=\frac{4n-5}{n-1}.$$ For $1<n<5/4$ the point $P_{rad}$ is a future attractor (not adequate to the present cosmological paradigm). Otherwise, $P_{rad}$ is a saddle equilibrium point.

\item The critical point $$P_{m-c}=\left(\frac{2n-3}{n},\frac{6(n-1)}{n}\right),\;$$ is associated with matter-curvature scaling. It is characterized by the following values of the relevant physical parameters
 
\bea &&\hat\Omega_m=-\frac{3-11 n+7 n^2}{n^2},\nonumber\\
&&q=-\frac{n-3}{n},\;\;w_{cur}=\frac{3(2n-1)}{8n-3},\nonumber\eea while the eigenvalues of the corresponding linearization matrix can be written, in compact form, as it follows: $$\lambda_{1,2}=\frac{3n^2-6n+3\pm\sqrt{s}}{2n(n-1)},$$ where $s\equiv 81+710 n^2-420 n+121 n^4-492 n^3$. The above point exists, i. e., it belongs in the phase space, whenever $0.35<n<1.3$. In other words, for $n$-s within the latter narrow interval, $\hat\Omega_m\geq 0$. Besides, for the allowed range of the parameter $n$, $q>0$ always, so that $P_{m-c}$ is associated with decelerated expansion. For $0.35<n<0.38$, and $1<n<1.22$, the equilibrium point $P_{m-c}$ is a saddle critical point, while for $0.38<n<1$, it is a stable spiral (future attractor). In the latter case, since $q>0$, the model is not suitable to accommodate the present cosmological paradigm.

\item The curvature-dominated critical point ($\hat\Omega_m=0$): $$P_{c-d}=\left(\frac{n(4n-5)}{(2n-1)(n-1)},-\frac{2(n-2)}{2n-1}\right),$$ is characterized by $$w_{cur}=-\frac{6n^2-7n-1}{3(2n^2-3n+1)},\;\;q=-\frac{2n^2-2n-1}{2n^2-3n+1},$$ and the following eigenvalues of the Jacobian matrix are obtained: $$\lambda_1=-\frac{2(7n^2-11n+3)}{2n^2-3n+1},\;\;\lambda_2=-\frac{4n-5}{n-1}.$$ The point $P_{c-d}$ can be associated with accelerated expansion for the following ranges of the free parameter $n$: $$\frac{1}{2}<n<1,\;\;\frac{1+\sqrt 3}{2}<n<\infty\;,$$ i. e., practically for all of the real segment (but for a very narrow interval). Meanwhile, the above eigenvalues are both simultaneously negative for the $n$-range: $5/4<n<\infty$. In other words, the (almost every where within the $n$-interval) inflationary equilibrium point $P_{c-d}$ is stable -- a future attractor -- but for the narrow interval $0\leq n<5/4$. For the particular value $n=2$, since $x=2\;\Rightarrow\;\dot H=0$, this equilibrium point is associated with a de Sitter solution.

\end{enumerate} Notice that, given a value of the variable $y=\bar y_i$ in a critical point $P_i=(\bar x_i,\bar y_i)$, then, due to the definition of this variable, one can write the curvature scalar $X$ as a function of the scale factor $$X(a)=M a^{-\bar y_i/(n-1)}+X_0\;,$$ where $M$ and $X_0$ are integration constants. Hence, for instance, for the critical point $P_{c-d}$, $$X(a)\propto a^{\frac{2(n-2)}{(2n-1)(n-1)}}\;.$$ This means that, for $$n<\frac{1}{2},\;\;1<n<2\;,$$ the scalar curvature decreases with the expansion of the universe, while, for $$\frac{1}{2}<n<1,\;\;n>2\;,$$ the curvature unboundedly grows as the expansion proceeds, thus mimicking phantom behavior as a mere curvature effect. 

In general, the above results coincide with those in \cite{amendola}, but for the point associated with the so called by the authors ``$\phi$ matter dominated epoch'' (their point C), which, in our study corresponds to the equilibrium point $P_{rad}$, where the non-linear curvature effects mimic radiation. Besides, several expressions for the matter-scaling solution -- solution B in \cite{amendola} -- do not coincide with ours. Even if the authors of Ref. \cite{amendola} chose a different phase space variable $x_2\equiv-f/6f'H^2$ (instead of our $x\equiv X/6H^2$), nonetheless, the results should coincide. It seems to us that the variable $x_2$ in \cite{amendola} is a bad choice after all. Actually, in the first stages of the study published in Ref. \cite{quiros}, we started by using $x_2$ as one of the phase space variables. Unfortunately, a problem arose: the limit in which the unmodified $f(X,Y,Z)$ is recovered from its DBI modification $\bar f(X,Y,Z)$ was not obtained! We were then forced to renounce to the variable $x_2$. This was, precisely, the motivation to use a different variable in \cite{quiros}, and in the present investigation as well. We recognize that this is a tricky situation, and that an independent investigation of the nature of this inconsistency is necessary, but this is behind the scope of the present paper.

\begin{table*}[ht!]\caption[crit]{Results of the numerical study of the eigenvalues of the linearization matrix corresponding to the equilibrium point $P_{tw}$ in subsection C of section V.}
\begin{tabular}{@{\hspace{4pt}}c@{\hspace{14pt}}c@{\hspace{14pt}}c@{\hspace{14pt}}
c@{\hspace{14pt}}c@{\hspace{14pt}}c@{\hspace{14pt}}c@{\hspace{14pt}}c@{\hspace{14pt}}
c@{\hspace{14pt}}c@{\hspace{14pt}}c@{\hspace{14pt}}c}
\hline\hline\\[-0.3cm]
$n$ &$x$&$y$&$z$&$\lambda_1$& $\lambda_2$&$\lambda_3$\\[0.1cm]\hline\\[-0.2cm]
$2$& $2$&$0$&$-1.44$&$-8.64$&$2.54$&$-2.89$&\\[0.2cm]
$2.5$& "&"&$-1.04$&$-3.6+4.5 i$ &$-3.6-4.5 i$&$-1.8$\\[0.2cm]
$5$& "&"&$25.5$&$-5.76$&$-3.95$&$0.7$\\[0.2cm]
$10$& "&"&$3144$&$-6$&$-4$&$1$\\[0.2cm]
\hline \hline\\[-0.3cm]
\end{tabular}\label{tab1}
\end{table*}

\begin{table*}[t!]\caption[crit]{Results of the numerical investigation corresponding to the equilibrium point $P_*$ in subsection C of section V. The values of the free parameter $n$ are approximately concentrated within the range $2<n<3$, since, for other values, the effective dimensionless energy density parameter $\hat\Omega_m$ is negative, meaning that the corresponding point does not belong in the phase space. The value $n=1$ is an isolated one. For $n=3$, $P_*$ is a non-hyperbolic critical point.}
\begin{tabular}{@{\hspace{4pt}}c@{\hspace{14pt}}c@{\hspace{14pt}}c@{\hspace{14pt}}
c@{\hspace{14pt}}c@{\hspace{14pt}}c@{\hspace{14pt}}c@{\hspace{14pt}}c@{\hspace{14pt}}
c@{\hspace{14pt}}c@{\hspace{14pt}}c@{\hspace{14pt}}c@{\hspace{14pt}}c@{\hspace{14pt}}c@{\hspace{14pt}}c}
\hline\hline\\[-0.3cm]
$n$ & $x$ & $y$ & $z$ & $\lambda_1$ & $\lambda_2$&$\lambda_3$& $\hat\Omega_m$ & $w_{cur}$ & $q$ \\[0.1cm]\hline\\[-0.2cm]
$1$ & $0$ & $1$ & $0$ & $6$ & $-1$ & $4$ & $0$ & $1/3$ & $1$ \\[0.2cm]
$2.02$ & $1.17$ & $0.02$ & " & $-3.2+13.5 i$ & $-3.2-13.5 i$ & $3.33$ & $0.27$ & $-0.98$ & $-0.2$ \\[0.2cm]
$2.05$ & $1.23$ & $0.04$ & " & $-3.3+9.62 i$ & $-3.3-9.62 i$ & $3.16$ & $0.24$ & $-0.97$ & $-0.23$ \\[0.2cm]
$2.1$ & $1.33$ & $0.07$ & " & $-3.4+6.9 i$ & $-3.4-6.9 i$ & $2.8$ & $0.2$ & $-0.95$ & $-0.32$ \\[0.2cm]
$2.3$ & $1.61$ & $0.11$ & " & $-3.8+3.86 i$ & $-3.8-3.86 i$ & $1.8$ & $0.1$ & $-0.93$ & $-0.6$ \\[0.2cm]
$2.5$ & $1.8$ & $0.1$ & " & $-4.1+2.7 i$ & $-4.1-2.7 i$ & $1.1$ & $0.05$ & $-0.95$ & $-0.8$ \\[0.2cm]
$2.7$ & $1.9$ & $0.07$ & " & $-4.3+2.01 i$ & $-4.3-2.01 i$ & $0.53$ & $0.02$ & $-0.97$ & $-0.9$ \\[0.2cm]
$2.9$ & $1.97$ & $0.24$ & " & $-4.4+1.55 i$ & $-4.4-1.55 i$ & $0.15$ & $0.005$ & $-0.99$ & $-0.97$ \\[0.2cm]
$3$ & $2$ & $0$ & " & $(-9+\sqrt 7 i)/2$ & $(-9-\sqrt 7 i)/2$ & $0$ & $0$ & $-1$ & $-1$ \\[0.2cm]
\hline \hline\\[-0.3cm]
\end{tabular}\label{tab2}
\end{table*}

\subsection{DBI Modification of the Dynamics}

In the present subsection we aim at investigating the way the above dynamical systems picture is affected by the Dirac-Born-Infeld strategy. Recall that, in order to consider the DBI modifications to the model in the above subsection ($f(X)=\mu X^n$), in the equations (\ref{psv}-\ref{general ode}) one has to replace $$f(X)\rightarrow\bar f(X)=\epsilon\lambda\left(\sqrt{1+\frac{2\epsilon f(X)}{\lambda}}-1\right)\;,$$ so that, for instance (compare with equations (\ref{psv}-\ref{general ode}))

\bea &&\hat\Omega_m\equiv\frac{\kappa^2\rho_m}{3\bar f'H^2}=1-x-y+\frac{\bar f}{6H^2 \bar f'},\nonumber\\
&&-\frac{\ddot{\bar f}'}{H^2 \bar f'}=2(x-2)+y+6\hat\Omega_m,\nonumber\\
&&w_{cur}=-1-\frac{2(x-2)+6\hat\Omega_m}{3\left[x-\frac{\bar f/\bar f'}{6H^2}+y\right]},\label{f c bar}\eea etc. 

The first consequence of the DBI modification of $f(X)=\mu X^n$, is that the system of ODE (\ref{general ode}) is not a closed one, so that, in addition to the variables $$x\equiv\frac{X}{6H^2},\;\;y\equiv-\frac{\dot{\bar f}'}{H \bar f'}$$ one has to consider the new phase space variable $$z\equiv\frac{\epsilon\lambda}{\mu (6H^2)^n}\;.$$ Hence, for instance $$\frac{\bar f}{6H^2 \bar f'}=\frac{z(\beta-1)\beta}{n x^{n-1}}\;,$$ where $\beta=\beta(x,y,z)\equiv\sqrt{1+2x^n/z}$. It is also verified that $$\frac{dX}{6H^2 d\tau}=-\left(\frac{z+2x^n}{(n-1)z+(n-2)x^n}\right)xy\;.$$ After appropriate algebraic manipulations one obtains the following autonomous system of ODE for the DBI modified version of the model $f(X)=\mu X^n$:

\bea &&\frac{dx}{d\tau}=-\left[\frac{z+2x^n}{(n-1)z+(n-2)x^n}\right]xy-2x(x-2),\nonumber\\
&&\frac{dy}{d\tau}=(x-2)(2-y)+y+y^2+6\hat\Omega_m,\nonumber\\
&&\frac{dz}{d\tau}=-2n(x-2)z,\label{dbi ode 1}\eea where now $$\hat\Omega_m=1-x-y+\frac{\beta(\beta-1)z}{nx^{n-1}}\;.$$ The corresponding (open) 3-dimensional phase space where to look for critical points of the above system of ODE, can be defined as $$\Psi=\{(x,y,z): x\geq 0,\;\hat\Omega_m\geq 0,\;2x^n+z\geq 0\}\;.$$ The latter bound comes from requiring that $\beta^2\geq 0$, while the former one $\hat\Omega_m\equiv\kappa^2\rho_m/6H^2\bar f'\geq 0$, is justified only if the free parameters ($\mu,n$), are both positive magnitudes. In fact, since $\bar f'=\beta^{-1}f'$, then, as long as $f'\geq 0$ (which is true only for positive ($\mu,n$) as noted above), hence $\bar f'$ is also non-negative. 

Notice that all of the equations and expressions in the unmodified case $f(X)=\mu X^n$ -- studied in the former subsection -- are recovered from the ones above, in the limit $z\rightarrow\infty$. Actually, in this limit: $$\lim_{z\rightarrow\infty}\frac{dX}{d\tau}=-\frac{xy}{n-1}\;,$$ while, since, up to second order in Taylor expansion $$\lim_{z\rightarrow\infty}\beta=1+\frac{x^n}{z}+{\cal O}(z^{-2})\;,$$ then $$\lim_{z\rightarrow\infty}\frac{\bar f}{6H^2\bar f'}=\frac{x}{n}\left(1+\frac{x^n}{z}\right)=\frac{x}{n}\;.$$ The relevant equilibrium points of the autonomous system of ODE (\ref{dbi ode 1}) are listed below, and their main properties are also shown.

\begin{enumerate}

\item The inflationary, curvature-dominated, de Sitter equilibrium point in the phase space $\Psi$ (see the definition above), $$P_{dS}=\left(2,-1,-2^{n+1}\right)\;,$$ is characterized by the following values of the parameters of observational interest: $$\hat\Omega_m=0,\;\;w_{cur}=-1,\;\;q=-1\;.$$ The eigenvalues of the Jacobian matrix corresponding to $P_{dS}$ are $$\lambda_1=-7,\;\;\lambda_2=-4,\;\;\lambda_3=-2\;,$$ so that it is always the late-time attractor, independent of the value of the free parameter $n$. In this case the effective ``curvature fluid'' behaves as vacuum energy -- properly, as a cosmological constant.

\item A twin -- also inflationary -- de Sitter state in the phase space, corresponds to the point $$P_{tw}=\left(2,0,\frac{2^n(n-4)+k+1}{2}\right)\;,$$ where $k\equiv\sqrt{1+n 2^{n+1}}$. As for its twin-solution above, for this equilibrium point one obtains that $$\hat\Omega_m=0,\;\;w_{cur}=-1,\;\;q=-1\;.$$ This time the magnitudes of the eigenvalues are given by extremely huge (and complex) expressions, so that, in their place, we decided to present the results of a numerical study for several values of the parameter $n$ instead. These results are shown in Tab. \ref{tab1}. As seen, $P_{tw}$ can be either a saddle critical point, or an stable spiral. In the latter case $P_{tw}$ rivals with its twin-solution $P_{dS}$: two late-time attractors co-exist, a feature also found in the study presented in \cite{quiros}. In the former case, due to its transient character, the solution $P_{tw}$ could serve as an alternative explanation to primordial (early-time) inflation.

\item There is another equilibrium point $$P_*=(\bar x,\bar y,0)\;,$$ where $\bar x=\bar x(n)$ and $\bar y=\bar y(n)$ are complicated functions of the parameter $n$. The corresponding expressions for $\hat\Omega_m=\hat\Omega_m(n)$, $w_{cur}=w_{cur}(n)$ and $q=q(n)$, are also bizarre huge expressions of the argument $n$. That is the reason why we present only a numerical investigation of the properties of this equilibrium point (the results are displayed in Tab. \ref{tab2}). In general, whenever it exists, $P_*$ is a spiral saddle point, but for $n=1$ (see the discussion below), and for $n=3$, where it is a non-hyperbolic point. For the particular, isolated, value of the free parameter $n=1$, one has $P_*=(0,1,0)$, and $\hat\Omega_m=0$ (curvature-dominated phase). Besides, since $w_{cur}=1/3$, the curvature behaves as a radiation fluid, favoring decelerated expansion ($q=1$). This time, since the eigenvalues of the Jacobian matrix are of different signs: $$\lambda_1=6,\;\;\lambda_2=-1,\;\;\lambda_3=4\;,$$ then, this phase represents a saddle equilibrium point in the phase space. For other values $n\pm\delta n$ in the vicinity of $n=1$, since $\hat\Omega_n$ is negative definite ($\hat\Omega_n<0$), then the point $P_*$ does not belong in the phase space $\Psi$ (see the definition above). The delicate situation with this equilibrium point is associated with the fact that, since $z=0$, then, either $\mu\rightarrow\infty$, or $H^2\rightarrow\infty$. While the former case is not of physical interest, the latter one leads to states of unboundedly large curvature (perhaps a singularity). As long as $P_*$ is a saddle point, which means, in turn, that the corresponding state can be only asymptotically approached by the system, this is not catastrophic.

\end{enumerate} In general, since at a given equilibrium point $P=(\bar x, \bar y, \bar z)$, $\bar y=-\dot{\bar f}'/H\bar f'$, then, one can write the following integral in quadratures: $$\int\frac{dX}{X}\left\{\frac{(n-1)\lambda+\epsilon (n-2)\mu X^n}{\lambda+2\epsilon\mu X^n}\right\}=-\bar y\ln a+C_0\;,$$ where $C_0$ is an integration constant. Once the above integral is computed one gets $a=a(X)$. Hence, by inverting the latter function (whenever the inverse exists), one can write the scalar curvature as a function of the scale factor $X=X(a)$.

Worth noticing that, while for the unmodified $f(X)=\mu X^n$-theory the case with $n=1$ is just Einstein-Hilbert gravity with only one critical point: the matter-dominated solution associated with decelerated expansion, for its DBI-dual $\bar f(X)$ there are found: i) the inflationary de Sitter late-time attractor (critical point $P_{dS}$), ii) the -- also inflationary -- de Sitter twin state (point $P_{tw}$); a saddle in the phase space, and, iii) the point $P_*$ -- associated with a decelerated expansion-phase -- where the non-linear curvature effects mimic radiation. The latter being also a saddle point in $\Psi$. 

Although for other values of the parameter $n$, the effects of the DBI modification are not so spectacular as for $n=1$, nonetheless, these are appreciable. In particular, i) the co-existence of two inflationary future attractors (equilibrium points $P_{dS}$ and $P_{tw}$) for given values of the free parameter $n$, and, ii) the chance to explain, in a united picture, primordial and late time inflation as originated by the non-linear effects of the curvature, otherwise.

\section{Discussion}

It is known since long ago, that quadratic (higher-order in general) modifications of general relativity are plagued by instabilities. Amongst them, we can name, the Ostrogradski, Ricci and tachyon instabilities, and the presence of Weyl ghosts -- also known as poltergeist, etc. To the list here we added the requirement of non-negativity of the effective gravitational coupling upon linearization, a subject that has not been much discussed in the literature, but, as we have shown in subsection A of section II, is of not less importance than the other kinds of catastrophic diseases of quadratic theories. Actually, linearization of an $f(X,....)$-theory around vacuum, maximally symmetric spaces of constant curvature $X_0$, leads to multiplication of the coupling constant $1/2\kappa^2$ by an overall factor $\alpha_0\equiv f_X^0-X_0f_{XX}^0$, which, in general, is not restricted to be positive. For the particular case of an $f(X)$ theory, Ricci stability, absence of tachyon instability, and non-negativity of the effective gravitational coupling, are not independent requirements. Hence, for instance, a Ricci stable $f(X)$-theory with positive effective gravitational coupling upon linearization, is also free of tachyon instability. Usually, for most $f(X,...)/f(X)$-theories found in the literature, not all of the above mentioned instabilities are surmounted at once. For instance, $f(X,Y,Z)$ theories where the invariants $Y$ and $Z$ enter in the combination $Z-4Y$ (like in the Gauss-Bonnet invariant), are poltergeist-free, while their linearization is also Ostrogradski stable. However, in general, these are either Ricci unstable, or have a scalar tachyon instability, or both. This is not to speak about positivity of the effective gravitational coupling upon linearization. Strictly speaking, most one can expect is to avoid really catastrophic instabilities such as the Ostrogradski and Ricci ones, and, after our results, also to allow for a positive definite effective gravitational coupling. Hence the question: does further modification of quadratic theories really matter?, in particular the one that gives name to the present paper: does Dirac-Born-Infeld modification of quadratic theories really matter? 

The results discussed in this paper suggest that, although the DBI deformation does not affect the Ostrogradski stability, other important instabilities such as the Ricci and tachyon ones are indeed surmounted after applying the DBI modification procedure to the original theory. Sometimes this is achieved at the cost of renouncing to the original motivation of the DBI strategy itself: to avoid singularities. Next one has to care about a by-product of the DBI deformation: its impact on the cosmic dynamics. It is for sure that the DBI modification also affects the dynamics, since the replacement $f\rightarrow\epsilon\lambda(\sqrt{1+2\epsilon f/\lambda}-1)$, inevitably affects the field equations. In terms of variables of the phase space this is quite clear: as a first visible effect, due to the introduction of a new energy scale $\lambda$, the DBI procedure increases the dimension of the phase space.\footnote{Although, for the example studied in this paper this is the case, in general, this is not always true. For instance, for the model studied in \cite{quiros} the dimension of the phase space is unchanged under the DBI replacement. However, even if the dimension is not affected, the topology of the phase space is in fact modified through, for instance, the occurrence of bifurcations in the space of parameters.} Hence, the question now is: would the effect of the DBI modification broaden the possibilities of a given model to do cosmology, or would it favor the contrary effect? To answer to this question, since finding exact solutions to the modified cosmological equations is bizarre difficult, one can rely on the tools of the dynamical systems. In this regard, knowledge of the equilibrium points in the phase space corresponding to a given cosmological model is a very important information since, independent on the initial conditions chosen, the orbits of the corresponding autonomous system of ODE will always evolve for some time in the neighborhood of these points. Besides, if the point were a stable attractor, independent of the initial conditions, the orbits will always be attracted towards it (either into the past or into the future). Going back to the original cosmological model, the existence of the equilibrium points can be correlated with generic cosmological solutions that might really decide the fate and/or the origin of the cosmic evolution. In a sense the knowledge of the asymptotic properties of a given cosmological model is more relevant than the knowledge of a particular analytic solution of the corresponding cosmological equations. While in the later case one might evolve the model from given initial data giving a concrete picture that can be tested against existing observational data, the knowledge of the asymptotic properties of the model gives the power to realize which will be the generic behavior of the model without solving the cosmological equations. In the dynamical systems language, for instance, a given particular solution of the Einstein's equations is just a single point in the phase space. Hence, phase space orbits show the way the model drives the cosmological evolution from one particular solution into another one. Equilibrium points in the phase space will correspond to solutions of the cosmological (Einstein's) equations that, in a sense, are preferred by the model, i. e., are generic. The lack of equilibrium points that could be correlated with a given analytic solution of the model, amounts to say that this solution is not generic, otherwise it can be attained under a very carefully arrangement of the initial conditions only.

As an example, to illustrate the spectacular effect of the DBI deformation procedure on the cosmic dynamics generated by arbitrary gravity theories, let us consider general relativity, i. e., we take the Lagrangian density $2\kappa^2{\cal L}_g=\sqrt{|g|} X$, which is a particular case of the theory $f(X)=\mu X^n$ studied in subsection B of section V. Before going into the details of the modification of the asymptotic properties of this theory, by the effects of its DBI deformation, we have to briefly discuss about the stability of the linearization of the theory, around vacuum, maximally symmetric spaces of constant curvature, since this topic is central both, to discuss about the physical content of the theory, and to study its weak field limit. To start with, while the massless graviton is the only propagating linearized degree of freedom in general relativity, the linearized gravitational spectrum of its DBI-dual $$\bar f(X)=\epsilon\lambda\left(\sqrt{1+\frac{2\epsilon X}{\lambda}}-1\right)\;,$$ consists of the massless graviton plus a spin-0 massive propagating mode. We know, however, that this is also an achievement of $f(X)$ modifications of general relativity, so that, this was, in fact, an expected effect. To see the advantage of DBI-$f(X)\equiv\bar f(X)$ over just $f(X)$, consider the simplest ``power-law'' modification of general relativity $f(X)=\mu X^n$ ($\mu,n\geq 0$). When linearized around vacuum, maximally symmetric spaces of constant curvature $X_0$, the mass squared of the spin-0 propagating mode is given by $$m_0^2=-\frac{X_0}{3}\left(\frac{n-2}{n-1}\right)\;,$$ while the Ricci stability requirement, and positivity of the effective gravitational coupling,  amount to $$f'_0-X_0 f''_0=-n(n-2)\mu X_0^{n-1}\geq 0\;,$$ and $$f''_0=n(n-1)\mu X_0^{n-2}\;,$$ respectively. One sees that: i) for $0<n<1$ the model is Ricci unstable, shows scalar tachyon instability, and is unphysical, due to negative effective gravitational coupling, ii) for $1<n<2$ the model is stable and physical, while iii) for $n>2$, it is Ricci stable, but develops scalar tachyon instability, and is unphysical due to negative coupling. Hence, the only physically meaningful range of the free parameter $n$ is $1<n<2$. For its dual DBI-$f(X)$, or $\bar f(X)$ theory, one can check that the mass squared of the scalar degree of freedom upon linearization can be written as follows: $$\bar m_0^2=-\frac{X_0}{3}\left(\frac{(n-2)\lambda+\epsilon (n-4)\mu X_0^n}{(n-1)\lambda+\epsilon (n-2)\mu X_0^n}\right)\;,$$ while Ricci stability and positivity of the effective gravitational coupling, are now given by $$(n-1)\lambda+\epsilon (n-2)\mu X_0^n\geq 0\;,$$ and $$(n-2)\lambda+\epsilon (n-4)\mu X_0^n\leq 0\;,$$ respectively, with the simultaneous requirement that $$2\epsilon\mu X_0^n+\lambda>0\;.$$ For the $\epsilon=+1$ DBI embedding, the latter bound is always satisfied, while for $\epsilon=-1$, it is satisfied whenever $X_0^n<\lambda/2\mu$. For $\epsilon=+1$, the $\bar f(X)$-theory is Ricci stable and physically meaningful (positive definite effective gravitational coupling) if, either  $$X_0^n\leq\left(\frac{n-1}{2-n}\right)\frac{\lambda}{\mu},\;\;1<n<2\;,$$ or $$X_0^n\geq\left(\frac{n-2}{4-n}\right)\frac{\lambda}{\mu},\;\;2<n<4\;.$$ Meanwhile, for $\epsilon=-1$, it is both, Ricci stable and physically meaningful if, either $$\left(\frac{1-n}{2-n}\right)\frac{\lambda}{\mu}\leq X_0^n\leq\left(\frac{2-n}{4-n}\right)\frac{\lambda}{\mu},\;\;0\leq n\leq 1\;,$$ or $$X_0^n\leq\left(\frac{2-n}{4-n}\right)\frac{\lambda}{\mu},\;\;1<n<2\;.$$ We see that, in both cases: $\epsilon=\pm 1$, the range of the parameter $n$ where the the theory is Ricci stable and physically meaningful (and consequently, also free of tachyon instability), is wider in the DBI-dual $\bar f(X)$ with respect to the $f(X)$ modification of general relativity. 

Now we are in position to discuss possible modifications of the cosmic dynamics by the DBI procedure. Going back to general relativity $f(X)=X$, the only equilibrium point in the phase space, is the one associated with matter dominance, where the peace of the cosmic expansion is decelerating. The phase space structure of its DBI-dual $\bar f(X)=\epsilon\lambda(\sqrt{1+2\epsilon X/\lambda}-1)$, is in fact, more complex and rich. Actually, according to the results presented in subsection C of section V, there are found 3 equilibrium points: i) the inflationary de Sitter attractor $P_{dS}=(2,-1,-4)$ ($\hat\Omega_m=0$, $w_{cur}=-1$, $q=-1$), ii) the ``twin'' inflationary (de Sitter) solution $P_{tw}=(2,0,-(5-\sqrt 5)/2)$, and iii) the ``radiation''-like critical point $P_{rad}=(0,1,0)$ ($\hat\Omega_m$, $w_{cur}=1/3$, $q=1$), associated with decelerated expansion of the universe. The last two points $P_{tw}$ and $P_{rad}$, are saddle equilibrium points in the phase space. An interesting feature of the modified $\bar f(X)$-theory, is associated with the existence of the inflationary (de Sitter) solutions $P_{dS}$, and $P_{tw}$. Both differ in their Hubble rates of expansion. Actually, since $\mu z=\epsilon\lambda(6H^2)^{-n}$, then, for $P_{dS}$ we have that $$H_0^2=-\frac{\epsilon\lambda}{24\mu}\;,$$ while, for $P_{tw}$: $$H_0^2=-\frac{\epsilon\lambda}{3(5-\sqrt 5)\mu}\;.$$ Notice that the above solutions exist only for the $\epsilon=-1$ DBI embedding. It is clear that, thanks to the existence of $P_{dS}$ and $P_{tw}$, there is room in the modified $\bar f(X)$-theory to explain, in a united way, primordial inflation (saddle equilibrium point $P_{tw}$), and the present stage of the cosmic speedup (late-time attractor $P_{dS}$). According to this picture, both inflationary stages have their origin in the non-linear curvature effects. The fact that the primordial inflation is associated, in this case, with a saddle equilibrium point, explains in a natural way the exit from inflation as due to the (in)stability properties of the corresponding solution. We have to underline, notwithstanding, that this is just a toy model that might not be capable of accommodating the existing amount of observations. In particular, this objection is obvious in what concerns to the formation of structure, since there are no equilibrium points that could be associated with matter dominance.

\section{Conclusions}

In this paper we have performed a thorough investigation of the effects of applying the Dirac-Born-Infeld strategy to quadratic modifications of gravity. We paid special attention to the subtleties associated with linearization of these theories, as well as of their DBI-duals, around vacuum, maximally symmetric spaces of constant curvature, an issue that is central to discuss about the weak field limit, as well as on the physical content of these theories. We have shown, in particular, that one aspect to take into consideration under the above linearization procedure, is the check of the positivity (non-negativity) of the effective gravitational coupling. A theory whose linearization fails to give a positive (effective) gravitational coupling, is devoid of physical meaning. For the particular case of $f(R)$ theories, Ricci stability, absence of tachyon instability, and positivity of the effective gravitational coupling upon linearization, are not independent requirements. It has been demonstrated, also, that the severe instabilities inherent in quadratic modifications of general relativity, can be surmounted -- at least smoothed out -- by considering their DBI-duals. This achievement is, in some cases, at the cost to renounce to the original motivation of the DBI deformation strategy: to remove singularities. 

An important focus of this paper has been the study of the modifications carried on the asymptotic properties of quadratic gravity by the DBI procedure. By applying very simple recipes of the dynamical systems, it has been demonstrated that the structure of the phase space is, in fact, modified by the DBI deformation strategy. The dimension of the phase space is increased, which means, in turn, a richer asymptotic structure. Even if, in many cases, the results are not spectacular, we think that considering DBI-duals of quadratic modifications of gravity can be an interesting arena where to look for alternative explanations of such important cosmic mysteries as primordial inflation and the present cosmic speedup.

\begin{acknowledgments}
This work was partly supported by PROMEP UGTO-CA-3, DAIP-UG, and
CONACYT under grant numbers 56946, and I0101/131/07 C-234/07, for the
Instituto Avanzado de Cosmolog\'ia (IAC) collaboration. 
\end{acknowledgments}

\section{Appendix: Dynamical Systems}

Here we include brief tips of how to apply the dynamical systems tools in general. In order to apply these tools one has to follow the steps enumerated below: 

\begin{enumerate}

\item To identify the phase space variables that allow writing the system of cosmological equations in the form of an autonomus system of ordinary differential equations (ODE), say:\footnote{There can be several different possible choices, however, not all of them allow for the minimum possible dimensionality of the phase space.}$$x_i=(x_1,x_2,...x_n)\;.$$

\item With the help of the chosen phase space variables, to build an autonomous system of ODE out of the original system of cosmological equations ($\tau$ is the time-ordering variable, not necessarily the cosmic time): $$\frac{dx_i}{d\tau}=f_i(x_1,x_2,...x_n)\;.$$ Notice that the RHS of these equations do not depend explicitly on $\tau$ (that is the reason why the system is called autonomous).

\item To identify the phase space spanned by the chosen variables $(x_1,x_2,...x_n)$, that is relevant to the cosmological model under study. This amounts, basically, to define the range of the phase space variables that is appropriate to the problem at hand: $$\Psi=\{(x_1,x_2,...x_n):\text{bounds on the}\;x_i\text{-s}\}\;.$$ 

\item Finding the equilibrium points of the autonomous system of ODE, amounts to solve the following system of algebraic equations on $(x_1,x_2,...x_n)$: $$f_i(x_1,x_2,...x_n)=0\;.$$

\item Next one linearly expands the equations of the autonomous system of ODE in the neighborhood of the equilibrium points $\bar p_k=p_k(\bar x_1,\bar x_2,...\bar x_n)$, $k=1,2,...m$:\footnote{In general the number of equilibrium points is different from the dimension of the phase space: $m\neq n$.} I. e., one replaces $x_i\rightarrow\bar x_i+e_i$, where $e_i$ are the small (linear) perturbations around the equilibrium points. Hence the system of ODE becomes a system of linear equations to determine the evolution of the $e_i$-s: $$\frac{de_i}{d\tau}=\bar f_i+\sum_{j=1}^n\left(\frac{\partial f_i}{\partial x_j}\right)_{\bar p}e_j+{\cal O}(e_i^2)\;,$$ otherwise, since $\bar f_i=f_i(\bar p)=0$, then $$\frac{de_i}{d\tau}=\sum_j^n[M(\bar p)_i^j]\;e_j+{\cal O}(e_i^2)\;,$$ where we have introduced the linearization or Jacobian matrix $[M^j_i]=\partial f_i/\partial x_j$.

\item The next step is to solve the secular equation to determine the eigenvalues $\lambda_i$ of the linearization matrix at the given equilibrium point $\bar p$: $$\det|M(\bar p)^j_i-\lambda\;U^j_i|=0\;,$$ where $[U^j_i]$ is the unit matrix. 

\item Once the eigenvalues of the linearization around a given equilibrium point $\bar p$ have been computed, the evolution of the perturbations is given by $$e_i(\tau)=\sum_j^n (e_0)_i^j\exp{(\lambda_j\tau)}\;,$$ where the amplitudes $(e_0)_i^j$ are constants of integration.

\end{enumerate} If all of the eigenvalues have negative real parts, the perturbations decay with $\tau$, i. e., the equilibrium point is stable against linear perturabtions. The corresponding equilibrium point is said to be a future attractor. If at least one of the eigenvalues has positive real part, the perturbations grow with $\tau$ so that these are not stable in the direction spanned by the given eigenvalue. Hence the point is said to be a saddle. The perturbations around a given equilibrium point are unstable, in other words the point is a past attractor (a source point in the phase space), if all of the eigenvalues have positive real parts. Points whose linearization is characterized by complex eigenvalues are said to be spiral equilibrium points, and are commonly associated with oscillatory behavior of the corresponding solution. If at least one of the eigenvalues has a vanishing real part, the equilibrium point is said to be non-hyperbolic. In the latter case, in general, and unless some of the non-vanishing real parts of the eigenvalues are of opposite sign, one can not give conclusive arguments on the stability of the equilibrium point. Other techniques have to be applied.


\begin{thebibliography}{MMHOT03}



\bibitem{udw} R. Utiyama, B. S. DeWitt, J. Math. Phys. {\bf 3} (1962) 608.

\bibitem{stelle1} K. S. Stelle, Phys. Rev. D {\bf 16} (1977) 953.

\bibitem{qstring} N. D. Birrell, P. C. W. Davies, "Quantum Fields in Curved Spacetime" (Cambridge University Press, Cambridge, 1982); I. L. Buchbinder, S. D. Odintsov, I. L. Shapiro, "Effective Actions in Quantum Gravity" (IOP Publishing, Bristol, 1992); G. A. Vilkovisky, Class. Quant. Grav. {\bf 9} (1992) 895.

\bibitem{odintsov1} S. Nojiri, S. D. Odintsov, hep-th/0601213.

\bibitem{frspeedup} S. Capozziello, Int. J. Mod. Phys. D {\bf 11} (2002) 483; D. N. Vollick, Phys. Rev. D {\bf 68} (2003) 063510 [astro-ph/0306630].

\bibitem{positive} S. M. Carroll, V. Duvvuri, M. Trodden, M. S. Turner, Phys. Rev. D {\bf 70} (2004) 043528 [astro-ph/0306438].

\bibitem{carroll} S. M. Carroll, A. De Felice, V. Duvvuri, D. A. Easson, M. Trodden, M. S. Turner, Phys. Rev. D {\bf 71} (2005) 063513 [astro-ph/0410031].

\bibitem{olmo2005} G. J. Olmo, 
Phys. Rev. Lett. {\bf 95} (2005) 261102; 
Phys. Rev. D {\bf 72} (2005) 083505.

\bibitem{olmoprd2007} G. J. Olmo, 
Phys. Rev. D {\bf 75} (2007) 023511.

\bibitem{odintsovrev} S. Nojiri, S. D. Odintsov, arXiv:0807.0685.

\bibitem{faraonirev} T. P. Sotiriou, V. Faraoni, 
arXiv:0805.1726.

\bibitem{faraoni2008} V. Faraoni, 
arXiv:0810.2602.

\bibitem{sotiriou2008} T. P. Sotiriou, 
arXiv:0810.5594.

\bibitem{DeFelice:2010aj}
  A.~De Felice and S.~Tsujikawa,
  arXiv:1002.4928 [gr-qc].

\bibitem{starobinsky} A. A. Starobinsky, Phys. Lett. B {\bf 91} (1980) 99.


\bibitem{stelle} K. Stelle, Gen. Rel. Grav. {\bf 9} (1978) 353-371.

\bibitem{ovrut} A. Hindawi, B. A. Ovrut, D. Waldram, Phys. Rev. D {\bf 53} (1996) 5583-5596 [hep-th/9509142]; Phys. Rev. D {\bf 53} (1996) 5597-5608 [hep-th/9509147].

\bibitem{solganik} A. Nunez, S. Solganik, Phys. Lett. B {\bf 608} (2005) 189.

\bibitem{navarro} I. Navarro, K. V. Acoleyen, JCAP {\bf 0603} (2006) 008 [gr-qc/0511045].

\bibitem{chiba} T. Chiba, JCAP {\bf 0503} (2005) 008 [gr-qc/0502070].

\bibitem{fiorini} F. Fiorini, R. Ferraro, Int. J. Mod. Phys. A {\bf 24} (2009) 1686-1689 [arXiv:0904.1767]; R. Ferraro, F. Fiorini, Phys. Rev. D {\bf 78} (2008) 124019 [arXiv:0812.1981]; Phys. Rev. D {\bf 75} (2007) 084031 [gr-qc/0610067].

\bibitem{dbiproc} S. Capozziello, S. Carloni, A. Troisi, astro-ph/0303041; S. M. Carroll, V. Duvvuri, M. Trodden, M. S. Turner, Phys. Rev. D {\bf 70} (2004) 043528 [astro-ph/0306438].

\bibitem{odi} S. Nojiri, S. D. Odintsov, Phys. Rev. D {\bf 68} (2003) 123512 [hep-th/0307288]; M. Abdalla, S. Nojiri, S. D. Odintsov, Class. Quant. Grav. {\bf 22} (2005) L35 [hep-th/0409177].

\bibitem{eddington} A. S. Eddington, ``The Mathematical Theory of Gravity" (Cambridge University Press, 1924).

\bibitem{deser} S. Deser, G. W. Gibbons, Class. Quant. Grav. {\bf 15} (1998) L35.

\bibitem{comelli} D. Comelli, 
Phys. Rev. D {\bf 72} (2005) 064018 [gr-qc/0505088].

\bibitem{woodard} R. P. Woodard, astro-ph/0601672.

\bibitem{kawasaki} A. D. Dolgov, M. Kawasaki, Phys. Lett. B {\bf 573} (2003) 1.

\bibitem{faraoni2006} V. Faraoni, Phys. Rev. D {\bf 74} (2006) 104017.

\bibitem{faraoni2007} V. Faraoni, Phys. Rev. D {\bf 75} (2007) 067302.

\bibitem{nojiri} S. Nojiri, S. D. Odintsov, Phys. Rev. D {\bf 68} (2003) 123512; Gen. Rel. Grav. {\bf 36} (2004) 1765.

\bibitem{amendola} L. Amendola, R. Gannouji, D. Polarski, S. Tsujikawa, Phys. Rev. D {\bf 75} (2007) 083504; L. Amendola, D. Polarski, S. Tsujikawa, Phys. Rev. Lett. {\bf 98} (2007) 131302; Int. J. Mod. Phys. D {\bf 16} (2007) 1555.

\bibitem{faulkner} T. Faulkner, M. Tegmark, E. F. Bunn, Y. Mao, Phys. Rev. D {\bf 76} (2007) 063505 [astro-ph/0612569].

\bibitem{quiros} R. Garcia-Salcedo, T. Gonzalez, C. Moreno,
  Y. Napoles, Y. Leyva, I. Quiros, arXiv:0912.5048.

\bibitem{Copeland:2010jt}
  E.~J.~Copeland, S.~Mizuno and M.~Shaeri,
  arXiv:1003.2881 [hep-th].

\bibitem{ehlers} J. Ehlers, P. Geren, R. K. Sachs, J. Math. Phys. {\bf 9} (1968) 1344. 



\end{thebibliography}
\end{document}